\documentclass[12pt, a4paper]{article}
\usepackage{cite}
\usepackage{amsmath,amssymb}
\input{colordvi.tex}
\usepackage{comment}
\usepackage{bm}
\usepackage{url}
\usepackage{ifpdf}
\ifpdf        
  \usepackage{graphicx, hyperref, xcolor}     
\else     
  \usepackage[dvipdfmx]{graphicx, hyperref, xcolor}     
 \fi
 \usepackage{subcaption}

\definecolor{rossoferrari}{HTML}{D9073D}
\definecolor{mediumblue}{HTML}{0000CD}
\hypersetup{
setpagesize=false,
bookmarksnumbered=true,%
bookmarksopen=true,%
colorlinks=true,%
linkcolor=rossoferrari,
urlcolor=mediumblue,
citecolor=mediumblue,
}

\begin{document}

\begin{titlepage}

\begin{center}

\hfill UT-19-12\\

\vskip .75in

{\Large \bf 
Novel Flavon Stabilization with \\ \vspace{2mm} Trimaximal Neutrino Mixing
}

\vskip .75in

{\large
So Chigusa$^{(a)}$, Shinta Kasuya$^{(b)}$ and Kazunori Nakayama$^{(a,c)}$
}

\vskip 0.25in

$^{(a)}${\em Department of Physics, Faculty of Science,\\
The University of Tokyo,  Bunkyo-ku, Tokyo 113-0033, Japan}\\[.3em]
$^{(b)}${\em Department of Mathematics and Physics,\\
Kanagawa University, Kanagawa 259-1293, Japan}\\[.3em]
$^{(c)}${\em Kavli Institute for the Physics and Mathematics of the Universe (WPI),\\
The University of Tokyo,  Kashiwa, Chiba 277-8583, Japan}
\end{center}
\vskip .5in

\begin{abstract}

We construct a supersymmetric $S_4$ flavor symmetry model with one of
the trimaximal neutrino mixing patterns, the so-called $\mathrm{TM}_1$,
by using the novel way to stabilize flavons, which we proposed recently.
The flavons are assumed to have tachyonic supersymmetry breaking mass
terms and stabilized by higher-dimensional terms in the potential. We
can obtain the desired alignment structure of the flavon vacuum
expectation values to realize neutrino masses and mixings consistent
with the current observations. This mechanism naturally avoids the
appearance of dangerous cosmological domain walls.  Although we study an
$S_4$ model in this paper, our mechanism is universal and can be applied
to many flavor models based on discrete flavor symmetry.

\end{abstract}

\end{titlepage}


\renewcommand{\thepage}{\arabic{page}}
\setcounter{page}{1}
\renewcommand{\thefootnote}{\#\arabic{footnote}}
\setcounter{footnote}{0}

\newpage


\section{Introduction}
\label{sec:Intro}

Discrete flavor symmetry is often introduced in order to naturally explain the observed patterns of neutrino 
masses and mixings~\cite{Altarelli:2010gt,Ishimori:2010au}. The discovery of the non-zero reactor mixing 
angle $\theta_{13}$~\cite{An:2012eh,Ahn:2012nd} had a great impact on the model building of the flavor 
symmetry and there are several ways to construct models consistent with observational 
data~\cite{King:2013eh,King:2014nza,King:2017guk,King:2019gif}.

One of the drawbacks of such models is that the spontaneous breakdown of the discrete 
flavor symmetry may lead to the formation of domain walls in the early universe, 
which is problematic in cosmology. Some of the discrete symmetry may be
anomalous under the gauge interaction~\cite{Preskill:1991kd,
Ibanez:1991pr, Ibanez:1991hv, Banks:1991xj, Araki:2007zza,
Araki:2008ek, Luhn:2008sa, Riva:2010jm, Chen:2013dpa, Chen:2015aba}
and thus softly broken by quantum effects, but it turns out that it
does not resolve the domain wall problem~\cite{Chigusa:2018hhl}. See
also Refs.~\cite{Antusch:2013toa,King:2018fke} for related works.

Recently we proposed a novel and simple way to avoid the domain wall problem in models with discrete 
flavor symmetry~\cite{Chigusa:2018yua}. The idea is that the flavons, which are Higgs fields responsible 
for the spontaneous breaking of flavor symmetry, are stabilized by higher dimensional potential balanced 
by tachyonic supersymmetry (SUSY) breaking mass terms. Then the flavor symmetry is already broken 
during inflation due to the negative Hubble-induced mass terms and never restored thereafter.\footnote{
In order for the flavon not to overshoot the origin of the scalar potential dynamically, it is essential to 
stabilize with higher dimensional potential terms~\cite{Chigusa:2018yua, Ema:2015dza}.
} Thus domain walls are inflated away and do not exist in the whole patch of the observable 
universe. Moreover, this novel flavon stabilization mechanism significantly simplifies the field content of 
the flavon sector. We do not need to introduce driving fields~\cite{Altarelli:2005yx} or any other additional 
field to obtain the flavon vacuum expectation values (VEVs) with desired alignment structure.\footnote{
See Refs.~\cite{Pascoli:2016eld,deMedeirosVarzielas:2017hen,deMedeirosVarzielas:2019hur} for 
stabilization of flavons in non-SUSY case.
} An explicit construction based on the $A_4$ flavor symmetry was made in Ref.~\cite{Chigusa:2018yua}.

In this paper, we apply the general argument of Ref.~\cite{Chigusa:2018yua} to the case of $S_4$ flavor 
symmetry. Models based on the $S_4$ flavor symmetry can lead to the so-called trimaximal neutrino 
mixings~\cite{Mohapatra:2003tw, Haba:2006dz, He:2006qd, Grimus:2008tt, Albright:2010ap, 
Ishimori:2010fs, He:2011gb, King:2011zj, Rodejohann:2012cf, Ding:2013hpa, Luhn:2013vna,
Shimizu:2017fgu}.  In particular, certain structure of the flavon VEVs may lead to one of the trimaximal 
patterns, the so-called TM$_1$, which is consistent with current observations~\cite{Luhn:2013vna,King:2019gif}.  
We will explicitly show that the flavon VEV alignment that leads to the TM$_1$ is realized along the line of 
Ref.~\cite{Chigusa:2018yua} and the resulting neutrino masses and mixings are consistent with the recent
observational data~\cite{Esteban:2018azc,nufit}.  This mixing pattern TM$_1$ will be tested in the near future 
by the more precise measurement of neutrino mixing angles.

In Sec.~\ref{sec:S4}, we briefly overview our setup of the $S_4$ flavor model. In Sec.~\ref{sec:neutrino}, 
we see that particular flavon VEV alignments lead to the so-called TM$_1$ pattern of the neutrino 
mixings and it is consistent with current observations. A novel way to obtain such a flavon VEV 
alignment is explained in detail in Sec.~\ref{sec:general} and we give a concrete example in 
Sec.~\ref{sec:flavon}. We conclude in Sec.~\ref{sec:conc}.

\section{Brief description of $S_4$ flavor model}
\label{sec:S4}

The superpotential consists of the charged lepton sector $W_\ell$, neutrino sector $W_\nu$ and 
flavon sector $W_{\rm f}$:
\begin{align}
	W = W_\ell + W_\nu + W_{\rm f}.
\end{align}
We focus on $W_\ell$ and $W_\nu$ in this section. The flavon part $W_{\rm f}$ will be discussed 
in detail in Sec.~\ref{sec:general} and Sec.~\ref{sec:flavon} where we explain how the desired 
flavon VEV alignments are obtained.

The superpotential of the charged lepton sector is assumed to be
\begin{align}
	W_\ell = \frac{y_\tau}{\Lambda} \tau^c H_d (\phi_{\ell} \ell)_{\bf 1}
	+ \frac{y_\mu}{\Lambda^2} \mu^c H_d (\phi_{\ell}^2 \ell)_{\bf 1'}
	+ \frac{1}{\Lambda^3} e^c H_d \left( y_e' \phi_{\ell}^3 \ell \right)_{\bf 1},
	\label{W_yukawa}
\end{align}
where $\ell$ is the lepton doublet, $e^c$, $\mu^c$, and $\tau^c$ are
respectively the right-handed electron, muon, and tau superfields, $H_d$
is the down-type Higgs doublet, $\phi_\ell$ is the flavon field in the
charged lepton sector, $\Lambda$ denotes the cutoff scale and
$y_e',y_\mu$ and $y_\tau$ are coupling constants. Charge assignments
under the $S_4$ flavor symmetry are summarized in Table~\ref{table:S4}.
See App.~\ref{app:S4} for our convention and notation of the $S_4$ group
representations and $S_4$ products.  There are three possible
contractions in the third term of \eqref{W_yukawa}, but all of them lead
to the same structure of the mass matrix in the argument below.  After
taking the VEV of
\begin{align}
	\left< \phi_{\ell} \right> = (0, v_\ell, 0)^T,\label{eq_vev_ell}
\end{align}
we obtain the diagonal charged lepton mass matrix as
\begin{align}
	\mathcal M_\ell = \frac{v_\ell v_d}{\Lambda} 
	\begin{pmatrix} y_e v_\ell^2/\Lambda^2 & 0 & 0 \\ 0 & 2y_\mu v_\ell/\Lambda & 0 \\ 0 & 0 & y_\tau \end{pmatrix},
\end{align}
with $v_d$ being the VEV of the down-type Higgs.  Note that $y_e$ in
this expression is some linear combination of $y_e'$s defined for each
contraction of the third term in \eqref{W_yukawa}.  One can take charged
lepton masses real and positive without loss of generality.  The mass
hierarchy of the charged leptons may be explained for $v_\ell / \Lambda
\sim \mathcal O(0.1)$. Since the charged lepton mass matrix is already
diagonal, we only have to consider the structure of the neutrino mass
matrix when discussing the lepton mixings in the weak interaction.

The superpotential of the neutrino sector is written as
\begin{align}
	W_\nu = \frac{H_u^2}{\Lambda^2}\left[ c_1\phi_{\bf 1}(\ell \ell)_{\bf 1} +c_2 \phi_{\bf 2} (\ell \ell)_{\bf 2}  
	+ c_{3'}  \phi_{\bf 3'} (\ell \ell)_{\bf 3'} +  c_{\psi} \psi_{\bf 3'} (\ell \ell)_{\bf 3'}  \right],
	\label{W_nu}
\end{align}
where $H_u$ denotes the up-type Higgs doublet. Here we have an
$S_4$-singlet $\phi_{\bf 1}$, an $S_4$-doublet $\phi_{\bf 2}$, and two
$S_4$-triplets $\phi_{\bf 3'}$ and $\psi_{\bf 3'}$.  Note that the
coupling of $\phi_\ell$ to the neutrino sector is forbidden by the
$Z_6^\ell$ symmetry whose charge assignments are given in
Table~\ref{table:S4}.  These flavons are assumed to develop VEVs of the
form
\begin{align}
	\left< \phi_{\bf 1} \right> = v_1,~~~
	\left< \phi_{\bf 2} \right> =  v_2 \begin{pmatrix} 1 \\ 1 \end{pmatrix},~~~
	\left< \phi_{\bf 3'} \right> =  v_{3'} \begin{pmatrix} 1 \\ 1 \\ 1 \end{pmatrix},~~~
	\left< \psi_{\bf 3'} \right> =   v_{\psi} \begin{pmatrix} 0 \\ 1 \\ -1 \end{pmatrix}.\label{eq_vev_nu}
\end{align}
Then the neutrino mass matrix 
is given by
\begin{align}
	\mathcal M_\nu = \frac{v_u^2}{\Lambda^2} \left[
	w_1\begin{pmatrix} 1 & 0 & 0 \\ 0& 0 & 1 \\ 0 & 1 & 0 \end{pmatrix}
	+w_2 \begin{pmatrix} 0 & 1 & 1 \\ 1& 1 & 0 \\ 1 & 0 & 1 \end{pmatrix}
	+w_{3'}\begin{pmatrix} 2 & -1 & -1 \\-1& 2 & -1 \\ -1 & -1 & 2 \end{pmatrix}
	+w_{\psi}\begin{pmatrix} 0 & 1 & -1 \\1& 2 & 0 \\ -1 & 0 & -2 \end{pmatrix} \right],
\end{align}
where $w_1 \equiv c_1 v_1$, $w_2 \equiv c_2 v_2$, $w_{3'} \equiv c_{3'} v_{3'}$ and 
$w_\psi \equiv c_{\psi} v_{\psi}$. Here $v_u$ is the VEV of the up-type Higgs.
For $v_{\psi}=0$, it becomes the neutrino mass matrix that is diagonalized by the 
tri-bimaximal mixing matrix~\cite{Harrison:2002er,Ma:2001dn}
\begin{align}
	U_{\rm TB} = \begin{pmatrix} 2/\sqrt{6} & 1/\sqrt{3} & 0 \\ -1/\sqrt{6}& 1/\sqrt{3} & 1/\sqrt{2} \\ -1/\sqrt{6} & 1/\sqrt{3} & -1/\sqrt{2} \end{pmatrix},
\end{align}
and the neutrino mass eigenvalues are given by
\begin{align}
	m_\nu = \frac{v_u^2}{\Lambda^2}\left(w_1-w_2 + 3w_{3'},~~w_1+2w_2,~~ -w_1+w_2 + 3w_{3'} \right).
\end{align}
The tri-bimaximal mixing, however, is already ruled out by experiments after the observation of 
non-zero $\theta_{13}$. In addition, the recent observation favors non-zero Dirac CP phase $\delta$.\footnote{
Neutrino mass eigenvalues obtained here are complex in general. One of them, $m_{\nu_1}$, for 
example, can be made real by the common phase rotation of $(\nu_1,\nu_2,\nu_3)$, which is 
accompanied by the opposite common phase rotation of $(e,\mu,\tau)$ to keep the mixing 
matrix intact. Thus there remain two physical Majorana phases. We always take such a basis in the following.
}

\begin{table}
\begin{center}
\begin{tabular}{|c|cccc|cc|c|cccc|} 
\hline
    ~        &  $\ell$ & $e^c$ & $\mu^c$ & $\tau^c$ & $H_u$ & $H_d$ & $\phi_{\ell}$ & $\phi_{\bf 1}$ & $\phi_{\bf 2}$ &$\phi_{\bf 3'}$ &$\psi_{\bf 3'}$  \\ \hline
 $S_4$     & ${\bf 3}$ & ${\bf 1}$ & ${\bf 1'}$ & ${\bf 1}$ & ${\bf 1}$ & ${\bf 1}$ & ${\bf 3}$& ${\bf 1}$& ${\bf 2}$& ${\bf 3'}$ & ${\bf 3'}$ \\ 
 U(1)$_R$& $5/6$ & $1/6$ & $1/2$ & $5/6$ & $0$ & $0$ & $1/3$& $1/3$& $1/3$& $1/3$ & $1/3$ \\ 
 $Z_6^\ell$& $0$ & $-3$ & $-2$ & $-1$ & $0$ & $0$ & $1$& $0$& $0$& $0$ & $0$ \\ 
 \hline
\end{tabular}
\caption{Charge assignments under $S_4$, $R$-symmetry U(1)$_R$ and $Z_6^\ell$ for leptons
and various Higgs and flavon fields.  \label{table:S4}
}
\end{center}
\end{table}

The non-zero $v_{\psi}$ breaks the tri-bimaximal symmetry but still there remains a $Z_2$ symmetry 
generated by the combination $SU$ of the $S_4$ group elements. It leads to the trimaximal mixing 
(especially, the so-called TM$_1$), which can fit the experimental results well~\cite{King:2019gif}.
It is convenient to divide the full mixing matrix $U^\nu$ into the tri-bimaximal part and the additional 
$2$-$3$ rotation part:
\begin{align}
	U^\nu = U_{\rm TB}U_{23},~~~~~~U_{23}\equiv  \begin{pmatrix}1 & 0 \\ 0 & u_{23} \end{pmatrix},
	\label{Unu}
\end{align}
where $u_{23}$ is a $2\times 2$ unitary matrix. Then we find
\begin{align}
	U^{\nu T} \mathcal M_\nu U^\nu 
	=\frac{v_u^2}{\Lambda^2}
	\begin{pmatrix} 
	w_1 -w_2 + 3w_{3'} & 0  \\ 
	0 & 
		u_{23}^T\begin{pmatrix} 
		w_1 +2w_2 & \sqrt{6} w_{\psi} \\ 
		 \sqrt{6} w_{\psi}  & -w_1+w_2 + 3w_{3'}
		\end{pmatrix} 
		u_{23}
	\end{pmatrix}.
	\label{UmU}
\end{align}
By choosing the unitary matrix $u_{23}$ appropriately, this can be diagonalized and we will 
obtain the full mixing matrix $U^\nu$ from which we can deduce the various neutrino mixing 
angles and CP phases. A concrete procedure will be discussed in the next section.

Assuming that all $w_1$, $w_2$, $w_{3'}$ and $w_\psi$ are the same order of magnitude and 
denoting their typical values by $w$, we obtain the neutrino mass scale as
\begin{align}
	m_\nu \sim 0.3\,{\rm eV} \left( \frac{10^{12}\,{\rm GeV}}{\Lambda} \right)^2 \left( \frac{w}{10^{10}\,{\rm GeV}} \right) \sin^2\beta,
\end{align}
where $\tan\beta \equiv v_u/v_d$. The typical value of the flavon VEV $w$ crucially depends on 
the stabilization mechanism. In our model presented in Sec.~\ref{sec:flavon}, it is given by 
$w\sim (m_{\rm soft} \Lambda^3)^{1/4}$ where $m_{\rm soft}$ is the soft SUSY breaking mass. 
Then we can reproduce the observed neutrino mass differences with some amount of tuning 
of $w_1,w_2,w_{3'}$ and $w_\psi$.

\section{Neutrino mixing in TM$_1$}
\label{sec:neutrino}

Let us describe how to diagonalize (\ref{UmU}) by choosing $u_{23}$. A general $2\times 2$ unitary 
matrix may be parameterized as
\begin{align}
	u_{23} = \begin{pmatrix} \cos\theta & e^{i\eta} \sin\theta \\ -e^{-i\eta} \sin\theta & \cos\theta \end{pmatrix}
	\begin{pmatrix} e^{i\alpha} & 0 \\ 0 & e^{i\beta} \end{pmatrix},
	\label{u23}
\end{align}
where $\theta$, $\eta$, $\alpha$ and $\beta$ are real parameters. 
As shown in App.~\ref{sec_diag}, we obtain the diagonal neutrino mass matrix by taking
\begin{align}
	\tan2\theta = \frac{2|C^*D + C B^*|}{|D|^2-|B|^2},~~~~~~e^{i\eta} = \frac{C^*D + C B^*}{|C^*D + C B^*|},
\end{align}
where $B= w_1+2w_2$, $C = \sqrt{6} w_{\psi}$ and $D=-w_1+w_2 + 3w_{3'}$, all of which may be complex 
in general. The parameters $\alpha$ and $\beta$ can be fixed if one wants to make the neutrino mass 
eigenvalues real so that the Majorana phases appear in the mixing matrix, although Majorana phases are 
irrelevant in the following discussion. Using those $\theta$ and $\eta$, we get the full mixing matrix as 
\begin{align}
	U^\nu =  \begin{pmatrix} 
	2/\sqrt{6} & \cos\theta/\sqrt{3} & e^{i\eta}\sin\theta/\sqrt{3} \\ 
	-1/\sqrt{6}& \cos\theta/\sqrt{3} - e^{-i\eta}\sin\theta/\sqrt{2} & \cos\theta/\sqrt{2} + e^{i\eta} \sin\theta/\sqrt{3} \\ 
	-1/\sqrt{6}& \cos\theta/\sqrt{3} + e^{-i\eta}\sin\theta/\sqrt{2} &-\cos\theta/\sqrt{2} + e^{i\eta} \sin\theta/\sqrt{3} 
	\end{pmatrix}
        \begin{pmatrix}
	 1 & 0 & 0 \\
	 0 & e^{i\alpha} & 0 \\
	 0 & 0 & e^{i\beta} 
	\end{pmatrix}.
\end{align}


On the other hand, the mixing matrix (MNS matrix) is usually parametrized as~\cite{Tanabashi:2018oca}
\begin{align}
	U^{\rm MNS}= \begin{pmatrix}
		c_{12}c_{13}    &    s_{12}c_{13}   &  s_{13}e^{-i\delta} \\
		-s_{12}c_{23}-c_{12}s_{23}s_{13}e^{i\delta} &  c_{12}c_{23}-s_{12}s_{23}s_{13}e^{i\delta} & s_{23}c_{13} \\
		s_{12}s_{23}-c_{12}c_{23}s_{13}e^{i\delta}  & -c_{12}s_{23}-s_{12}c_{23}s_{13}e^{i\delta}  &c_{23}c_{13} 
	\end{pmatrix}
	\begin{pmatrix} 1 & 0 & 0 \\ 0 & e^{i\alpha_{21}/2} & 0 \\ 0 & 0 & e^{i\alpha_{31}/2} \end{pmatrix},
	\label{U_MNS}
\end{align}
where $c_{ij} = \cos\theta_{ij}$, $s_{ij} = \sin \theta_{ij}$, $\delta$ is the Dirac CP phase, and 
$\alpha_{21}$ and $\alpha_{31}$ are Majorana CP phases. Here we take the basis in which the 
neutrino mass matrix is real and diagonal.

The mixing matrix $U^\nu$ can be reduced to the form of $U^{\rm MNS}$ (\ref{U_MNS}) by using the 
freedom to rotate the phase of charged leptons and neutrinos. Even without performing such a concrete phase 
rotation, one can conveniently deduce the mixing parameters and the Dirac CP phase by using the rotation 
invariant quantity as
\begin{align}
	s_{13}^2 = |U^\nu_{e3}|^2,~~~s_{12}^2 = \frac{|U^\nu_{e2}|^2}{1-|U^\nu_{e3}|^2},~~~s_{23}^2 =  \frac{|U^\nu_{\mu 3}|^2}{1-|U^\nu_{e3}|^2},
\end{align}
and
\begin{align}
	J_{\rm CP} \equiv {\rm Im}\left( U^\nu_{\mu 3}U^{\nu *}_{e3} U^\nu_{e2} U^{\nu *}_{\mu 2}\right) = s_{12}s_{23}s_{13} c_{12} c_{23} c_{13}^2\sin\delta.
\end{align}
We can also derive sum rules independent of $\theta$ and $\eta$. One of them is obtained from 
$|U^\nu_{e1}|^2=(c_{12}c_{13})^2$ as
\begin{align}
	c_{12}c_{13} = \frac{2}{\sqrt{6}} \simeq 0.816,    \label{c12c13}
\end{align}
which fits well with the observation: $c_{12}^{\rm (obs)} c_{13}^{\rm (obs)} = 0.821_{-0.008}^{+0.007}$ 
within $1\sigma$~\cite{Esteban:2018azc}. Another one is extracted from comparing $|U^\nu_{\mu1}|^2$ 
and $|U^\nu_{\tau1}|^2$ with the MNS parameterization:
\begin{align}
	\cos(2\theta_{23})\left(\frac{2}{3}-\cos(2\theta_{12})\right) + \sin(2\theta_{12})\sin(2\theta_{23}) \sin\theta_{13} \cos\delta=0.
\end{align}
Substituting the best-fit values $\theta_{23}^{\rm (obs)} = 49.6^\circ$, $\theta_{12}^{\rm (obs)} = 33.82^\circ$ and $\theta_{13}^{\rm (obs)}=8.61^\circ$ ~\cite{Esteban:2018azc}, we obtain $\cos\delta \simeq 0.33$.\footnote{
If one assumes that the deviation from the tri-bimaximal form is small, one may regard $\theta_{13}$ as a 
small parameter and expand as $\theta_{23} = \pi/4 + \Delta\theta_{23}$. Note that the deviation of 
$\theta_{12}$ from its tri-bimaximal value $(c_{12}=\sqrt{2/3})$ is second order in terms of $\theta_{13}$ 
because of the constraint (\ref{c12c13}). Then we have
\begin{align}
	\Delta \theta_{23} \simeq \sqrt{2} \theta_{13} \cos\delta.
\end{align}
}
Note that there are no more non-trivial sum rules in the present model. We have $4$ complex model 
parameters $w_1$, $w_2$, $w_{3'}$ and $w_\psi$ to construct neutrino mass matrix, but an overall 
phase is irrelevant and hence we have $7$ real parameters (2 of which are parameterized by 
$\theta$ and $\eta$). The number of the physical observables is $9$, consisting of three neutrino mass 
eigenvalues, three mixing angles, one Dirac CP phase, and two Majorana CP phases. Thus there 
should be two non-trivial sum rules in the physical observables.

\begin{figure}
\begin{center}
\begin{tabular}{cc}
\includegraphics[scale=1.1]{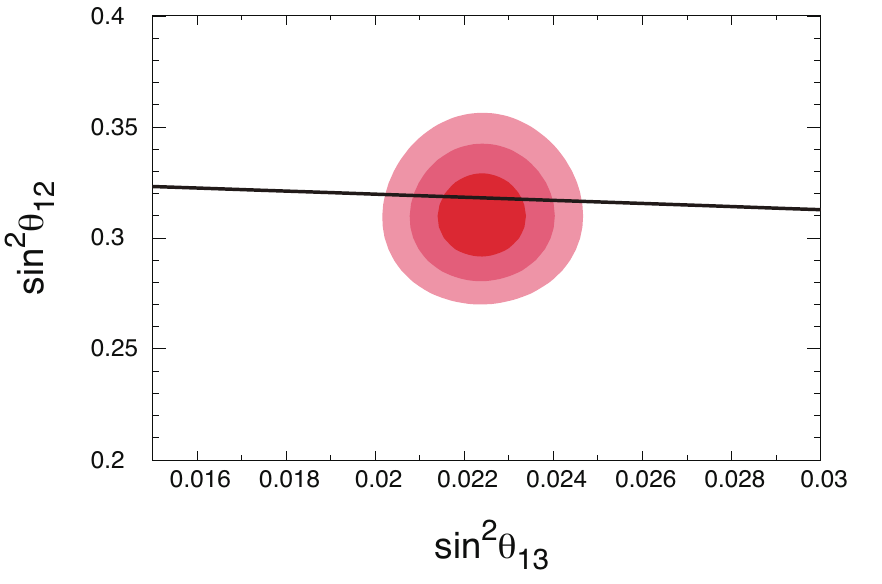} \\~\\
\includegraphics[scale=1.1]{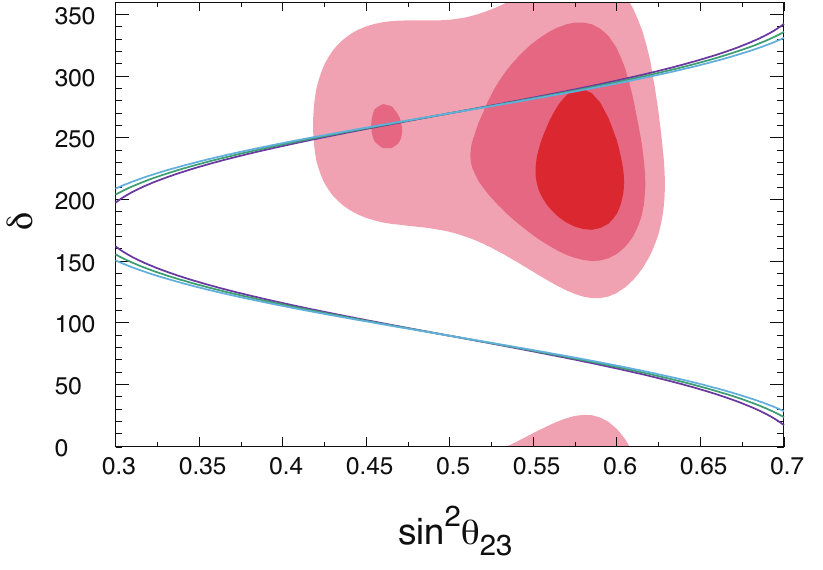}
\end{tabular}
\end{center}
\caption{
	(Top) Observationally preferred region of $\sin^2\theta_{13}$ and $\sin^2\theta_{12}$ within 
	$1\sigma$, $2\sigma$ and $3\sigma$ ranges~\cite{Esteban:2018azc,nufit}. A line indicates the 
	sum rule (\ref{c12c13}), a prediction of TM$_1$. 
	(Bottom) Observationally preferred region of $\sin^2\theta_{23}$ and $\delta$ within 
	$1\sigma$, $2\sigma$ and $3\sigma$ ranges~\cite{Esteban:2018azc,nufit}. Three lines indicate 
	the predictions of TM$_1$ for typical values of $\theta$ which reproduce the observed values of 
	$\theta_{12}$ and $\theta_{13}$.
}
\label{fig:TM1}
\end{figure}

Fig.~\ref{fig:TM1} shows observationally preferred region of $\sin^2\theta_{13}$ and 
$\sin^2\theta_{12}$ (top), and $\sin^2\theta_{23}$ and $\delta$ (bottom) within $1\sigma$,
$2\sigma$ and $3\sigma$ ranges~\cite{Esteban:2018azc,nufit}. A line in the top panel 
indicates the sum rule (\ref{c12c13}), a prediction of TM$_1$. This is solely determined 
by the model parameter $\theta$ and independent of $\eta$. It is seen that the line crosses
observationally preferred region and hence $\theta$ is constrained to some narrow range.  
Once we fix $\theta$, we can draw a line on the plane of $\sin^2\theta_{23}$ and $\delta$ 
by changing $\eta$. Three lines in the bottom panel indicate the predictions of TM$_1$ for
typical values of $\theta$ which reproduce the observed values of $\theta_{12}$ and 
$\theta_{13}$.  We can see that the TM$_1$ prediction is consistent with current experimental 
results. If future observations determine $\delta$ and $\theta_{23}$ more precisely, the
TM$_1$ can be tested more crucially.

\section{Novel flavon stabilization: general argument}
\label{sec:general}

Let us adopt the idea of Ref.~\cite{Chigusa:2018yua} to obtain the desired alignment structure 
of the flavon VEVs in the $S_4$ flavor model introduced in the previous sections.  In this section, we
describe some general arguments about our method for the flavon stabilization, while we will give a 
concrete example of the superpotential for the flavon sector in the next section.

The basic idea is that the flavon fields are stabilized by the balance between the negative soft SUSY 
breaking mass and non-renormalizable terms in the potential. The latter comes from non-renormalizable 
superpotentials. In our setup, as shown in Table~\ref{table:S4}, flavons have U(1)$_R$ charge of $1/3$ 
and hence only terms with the sixth power of flavons are allowed in the superpotential. Schematically, 
we have
\begin{align}
	W_{\rm f} \sim \frac{\phi^6}{\Lambda^3},
\end{align}
where the flavon fields are collectively denoted as $\phi$. This is a shorthand notation and there are 
actually many ways of contractions of $S_4$-charged flavons. Together with the SUSY breaking 
mass term, the scalar potential has the form as
\begin{align}
	V \sim -m^2|\phi|^2 + \frac{|\phi|^{10}}{\Lambda^6},
\end{align}
and the flavon fields obtain VEVs of $\left<|\phi|\right>\sim (m \Lambda^3)^{1/4}$. What is non-trivial is 
whether or not we can obtain the alignment structure given in \eqref{eq_vev_ell} and \eqref{eq_vev_nu}. 
Here we show that the configuration \eqref{eq_vev_ell} and \eqref{eq_vev_nu} is always an extremum 
of the scalar potential independent of the detailed form of the superpotential under some assumptions.

First let us consider $\phi_\ell$. Due to the $Z_6^\ell$ symmetry, it is not mixed with other flavons in the 
superpotential. Therefore, in terms of components $\phi_\ell=(\phi_{\ell,1},\phi_{\ell,2},\phi_{\ell,3})^T$, 
the superpotential may be generically written as
\begin{align}
	W_{\rm f,\ell} = \frac{1}{\Lambda^3}\left( \phi_{\ell,2}^6 + \phi_{\ell,2}^5 \phi_{\ell,1} + \phi_{\ell,2}^5 \phi_{\ell,3} + \cdots \right),
\end{align}
where dots represent terms with higher powers of $\phi_{\ell,1}$ or $\phi_{\ell,3}$ and we omitted 
$\mathcal O(1)$ numerical coefficients for simplicity.  First, by substituting $\phi_\ell=(0,\phi_{\ell,2},0)^T$ 
one can minimize the potential along the $\phi_{\ell,2}$ direction to find VEV $\left<|\phi_{\ell,2}|\right>=v_\ell$. 
A necessary and sufficient condition for this configuration to be an extremum of the potential is that 
there are no linear terms with respect to $\phi_{\ell,1}$ and $\phi_{\ell,3}$ in the superpotential 
when expanded around the configuration.\footnote{ 
Here ``linear terms with respect to $\phi_{\ell,i}$'' means terms of the form of $v^5 \phi_{\ell,i}/\Lambda^3$, 
where $v$ collectively represents the non-zero VEV of various flavons.  We will use the same terminology 
in the following discussion.  
}  Actually, we can show that such terms are forbidden by symmetry.  In our setup, the flavon 
superpotential must be invariant under $S_4$ symmetry and also $Z_6^\ell$ symmetry under 
which the flavon is rotated like $\phi_\ell \to \Omega \phi_\ell$ with $\Omega \equiv e^{2\pi i/6}$.  
Then, for example, the superpotential must be invariant under the transformation
\begin{align}
	\phi_\ell \to \Omega^2 T \phi_\ell = \begin{pmatrix}
		\Omega^2 & 0 & 0 \\ 0 & 1 & 0 \\ 0 & 0 & \Omega^4
	\end{pmatrix} \phi_\ell,
\end{align}
where $T$ is an element of three dimensional representation of $S_4$ given in (\ref{SUT3}). 
Under this transformation, $\phi_{\ell,2}$ is invariant but $\phi_{\ell,1}$ and $\phi_{\ell,3}$ are 
not. Therefore, we cannot have terms linear in $\phi_{\ell,1}$ and $\phi_{\ell,3}$ in the 
superpotential and the configuration \eqref{eq_vev_ell} is indeed an extremum of the potential. 
Note that the combination $\Omega^2 T$ used here is the generator of the remaining 
$Z_3 \subset S_4 \times Z_6^{\ell}$ symmetry that is retained in $W_{\mathrm{f},\ell}$ even 
after $\phi_\ell$ takes the non-zero VEV of \eqref{eq_vev_ell}. In order to show that it is a 
minimum of the potential, one must check terms quadratic in $\phi_{\ell,1}$ and $\phi_{\ell,3}$,
which depends on the detailed forms of the superpotential. In the next section, we will give 
a concrete example in which the configuration \eqref{eq_vev_ell} is actually a minimum of the potential.

Next, flavons $\phi_{\bf 1}$, $\phi_{\bf 2}$, $\phi_{\bf 3'}$ and $\psi_{\bf 3'}$ must have the same 
quantum numbers (except for the $S_4$ charge) due to the coupling (\ref{W_nu}) and hence they 
are in general mixed with each other in the flavon superpotential. What is dangerous is the mixing 
between $\phi_{\bf 3'}$ and $\psi_{\bf 3'}$ that would spoil the vacuum alignment \eqref{eq_vev_nu}.
In order to forbid dangerous mixings between $\phi_{\bf 3'}$ and $\psi_{\bf 3'}$, we want to assign 
different charges on them under some symmetry. For this purpose, one can modify the superpotential 
of the neutrino sector (\ref{W_nu}) as follows~\cite{Luhn:2013vna}:
\begin{align}
	W_\nu = \frac{H_u^2}{\Lambda^3}\left[ \xi_{\bf 1}\left( c_1\phi_{\bf 1}(\ell \ell)_{\bf 1} +c_2 \phi_{\bf 2} (\ell \ell)_{\bf 2}  
	+ c_{3'}  \phi_{\bf 3'} (\ell \ell)_{\bf 3'} \right)+  c_{\psi} \xi_{\bf 1}' \psi_{\bf 3'} (\ell \ell)_{\bf 3'}  \right],
\end{align}
where we introduced additional $S_4$ singlet flavons $\xi_{\bf 1}$ and $\xi'_{\bf 1}$ and imposed 
additional $Z_6^\xi$, $Z_6^{\xi'}$ and $Z_{12}^{\xi\xi'}$ symmetry as presented in Table~\ref{table:flavon}. 
Note that U(1)$_R$ charge assignments are modified from those of Table~\ref{table:S4}. These additional 
flavons are assumed to have VEVs of $\left<\xi_{\bf 1}\right>=v_\xi$ and $\left<\xi'_{\bf 1}\right>=v_{\xi'}$. 
The structure of neutrino masses and mixings described in Sec.~\ref{sec:S4} and \ref{sec:neutrino} are 
unchanged after reinterpreting $w_1 = c_1v_{\xi} v_1/\Lambda$, and so on. The additional $Z_6^\xi$, 
$Z_6^{\xi'}$ and $Z_{12}^{\xi\xi'}$ symmetry, combined with U(1)$_R$ symmetry, restrict the form 
of the flavon superpotential to be
\begin{align}
	W_{\rm f} = W_{\rm f,\ell}(\phi_\ell)
        +W_{\rm f,\phi}(\phi_{\bf 1},\phi_{\bf 2},\phi_{\bf 3'})
        +W_{\rm f,\xi}(\xi_{\bf 1})
        +W_{\rm f,\psi}(\psi_{\bf 3'})
        +W_{\rm f,\xi'}(\xi'_{\bf 1}).
\end{align}
It is easy to see that $\phi_{\bf 1}$, $\xi_{\bf 1}$ and $\xi'_{\bf 1}$ are stabilized by using terms like 
$\phi^6 / \Lambda^3$, where $\phi$ collectively denotes a $S_4$ singlet flavon field. Thus we only 
need to focus on $\phi_{\bf 2}, \phi_{\bf 3'}$ and $\psi_{\bf 3'}$ below.

\begin{table}
\begin{center}
\begin{tabular}{|c|cccc|cc|c|cccc|cc|} 
\hline
    ~        &  $\ell$ & $e^c$ & $\mu^c$ & $\tau^c$ & $H_u$ & $H_d$ & $\phi_{\ell}$ & $\phi_{\bf 1}$ & $\phi_{\bf 2}$ &$\phi_{\bf 3'}$ &$\psi_{\bf 3'}$ & $\xi_{\bf 1}$ & $\xi'_{\bf 1}$  \\ \hline
 $S_4$     & ${\bf 3}$ & ${\bf 1}$ & ${\bf 1'}$ & ${\bf 1}$ & ${\bf 1}$ & ${\bf 1}$ & ${\bf 3}$& ${\bf 1}$& ${\bf 2}$& ${\bf 3'}$ & ${\bf 3'}$ & ${\bf 1}$ & ${\bf 1}$ \\ 
 U(1)$_R$& $2/3$ & $1/3$ & $2/3$ & $1$ & $0$ & $0$ & $1/3$& $1/3$& $1/3$& $1/3$ & $1/3$ & $1/3$ & $1/3$\\ 
 $Z_6^\ell$& $0$ & $-3$ & $-2$ & $-1$ & $0$ & $0$ & $1$& $0$& $0$& $0$ & $0$ & $0$ & $0$\\
 $Z_6^\xi$   & $0$ & $0$ & $0$ & $0$ & $0$ & $0$ & $0$& $-1$& $-1$& $-1$ & $0$ & $1$ & $0$\\
 $Z_6^{\xi'}$& $0$ & $0$ & $0$ & $0$ & $0$ & $0$ & $0$& $0$& $0$& $0$ & $-1$ & $0$ & $1$\\
 $Z_{12}^{\xi\xi'}$ & $5$ & $-5$ & $-5$ & $-5$ & $0$ & $0$ & $0$ & $0$ & $0$ & $0$ & $0$ & $2$ & $2$\\ 
 \hline
\end{tabular}
\caption{Charge assignments under $S_4$, $R$-symmetry U(1)$_R$, $Z_6^\ell$, $Z_6^\xi$, $Z_6^{\xi'}$ 
and $Z_{12}^{\xi\xi'}$ for leptons, Higgs and flavon fields.  \label{table:flavon}
}
\end{center}
\end{table}

For $\psi_{\bf 3'}$, it is convenient to work with the basis in which $\overline \psi_{\bf 3'} = V_{\psi} \psi_{\bf 3'}$, where
\begin{align}
	V_\psi=\begin{pmatrix}
		1 & 0 & 0 \\ 0 & 1/\sqrt 2 & -1/\sqrt 2 \\ 0 & 1/\sqrt 2 & 1/\sqrt 2
	\end{pmatrix}.
\end{align}
In this basis, the alignment structure \eqref{eq_vev_nu} becomes 
$\left< \overline\psi_{\bf 3'} \right> =  \sqrt 2 v_{\psi} (0,1,0)^T$. By noting that $\overline\psi_{{\bf 3}'}$ transforms 
as $\overline\psi_{{\bf 3}'} \to V_{\psi} X V_{\psi}^{-1} \overline\psi_{\bf 3'}$ under any $S_4$ group element 
$X$, one finds that $\overline\psi_{{\bf 3'},2}$ is invariant but $\overline\psi_{{\bf 3'},1}$ and 
$\overline\psi_{{\bf 3'},3}$ are not under the transformation
\begin{align}
	\Omega^3 \,V_\psi U V_\psi^{-1} = \begin{pmatrix}
		  -1 & 0 & 0 \\ 0 & 1 & 0 \\ 0 & 0 & -1 
	\end{pmatrix}.
\end{align}
Here $\Omega^3$ should be regarded as an element of $Z_6^{\xi'}$ rotation. Thus we cannot have terms 
linear in $\overline\psi_{{\bf 3'},1}$ and $\overline\psi_{{\bf 3'},3}$ and hence the $\psi_{\bf 3'}$ configuration 
in \eqref{eq_vev_nu} is always an extremum of the potential.  In this case, $\Omega^3 S$ and $\Omega^3 U$ 
are the generators of the $Z_2 \times Z_2 \subset S_4 \times Z_6^{\xi'}$ subgroup that is unbroken in 
$W_{\mathrm{f},\psi}$ when the VEV of $\psi_{\bm{3'}}$ in \eqref{eq_vev_nu} is taken into account.  In addition, 
since $\xi_{\bm{1}}'$ also acquires the non-zero VEV, the symmetry in the neutrino sector is further broken to 
the $Z_2$ subgroup generated by $(\Omega^3 S)(\Omega^3 U) = SU$, which is responsible for the 
$\mathrm{TM}_1$ mixing patterns.

Similar arguments hold for the other flavons, $\phi_{\bf 2}$ and $\phi_{\bf 3'}$. In general, they are mixed 
in the superpotential $W_{\rm f,\phi}(\phi_{\bf 1},\phi_{\bf 2},\phi_{\bf 3'})$.  Let us go to the basis in which 
$\overline \phi_{\bf 2} = V_{\bf 2} \phi_{\bf 2}$ and $\overline \phi_{\bf 3'} = V_{\phi} \phi_{\bf 3'}$ where
\begin{align}
	V_{\bf 2} =\frac{1}{\sqrt 2}\begin{pmatrix}
		  1 & 1 \\ -1 & 1 
	\end{pmatrix},~~~
	V_\phi = \begin{pmatrix}
		2/\sqrt{6} & -1/\sqrt 6 & -1/\sqrt 6 \\ 1/\sqrt 3 & 1/\sqrt3 & 1/\sqrt 3 \\ 0 & 1/\sqrt 2 & -1/\sqrt 2
	\end{pmatrix},
\end{align}
In this basis, the alignment structure \eqref{eq_vev_nu} becomes
\begin{align}
	\left< \overline\phi_{\bf 2} \right> =  \sqrt 2 v_{\bf 2} \begin{pmatrix} 1 \\ 0 \end{pmatrix},~~~
	\left< \overline\phi_{\bf 3'} \right> = \sqrt 3 v_{\bf 3'} \begin{pmatrix} 0 \\ 1 \\ 0 \end{pmatrix}.
\end{align}
Let us note that
\begin{align}
	V_{\bf 2} S V_{\bf 2}^{-1} = \begin{pmatrix}
		  1 & 0 \\ 0 & 1 
	\end{pmatrix},~~~
	V_{\bf 2} SU V_{\bf 2}^{-1} = \begin{pmatrix}
		  1 & 0 \\ 0 & -1 
	\end{pmatrix},
\end{align}
and
\begin{align}
	V_\phi S V_\phi^{-1} = \begin{pmatrix}
		  -1 & 0 & 0 \\ 0 & 1 & 0 \\ 0 & 0 & -1 
	\end{pmatrix},~~~
	V_\phi SU V_\phi^{-1} = \begin{pmatrix}
		  -1 & 0 & 0 \\ 0 & 1 & 0 \\ 0 & 0 & 1 
	\end{pmatrix}.
\end{align}
Therefore, from the invariance of the superpotential under the $SU$ transformation, terms linear 
in $\overline\phi_{{\bf 2},2}$ are forbidden.  
Likewise, from the invariance under the $S$ transformation, we cannot have terms linear in 
$\overline\phi_{{\bf 3}',1}$ and $\overline\phi_{{\bf 3}',3}$.
Therefore, for general superpotential $W_{\rm f,\phi}(\phi_{\bf 1},\phi_{\bf 2},\phi_{\bf 3'})$, 
we show that the configuration given in \eqref{eq_vev_nu} 
is always an extremum of the potential.  Again note that $S$ and $U$ are the generators of the
remaining $Z_2 \times Z_2 \subset S_4 \times Z_6^\xi$ symmetry of 
$W_{\mathrm{f},\phi}+W_{\mathrm{f},\xi}$ (which of course contains $SU$) when $\phi_{\bm{1}}$, 
$\phi_{\bm{2}}$, $\phi_{\bm{3'}}$ and $\xi$ obtain non-zero VEVs.  In order to show that it is in fact a
minimum of the potential, one must check the curvature around it. It depends on the detailed form 
of the superpotential. We will give a concrete example in which the configuration \eqref{eq_vev_ell} and
\eqref{eq_vev_nu} is actually a minimum of the potential.

\section{Novel flavon stabilization: concrete example}
\label{sec:flavon}

We have shown that the flavon configuration given by \eqref{eq_vev_ell} and \eqref{eq_vev_nu} is 
always an extremum of the potential by just using general symmetry argument in the previous section. 
Now we want to show that this configuration can actually be a minimum of the potential.  As a concrete 
example, let us consider a simple non-renormalizable superpotential of the flavon sector given as
\begin{align}
 W_{\mathrm{f}} &= W_{\mathrm{f}, \ell} + W_{\mathrm{f}, \bm{1}} +
 W_{\mathrm{f}, \bm{2}} + W_{\mathrm{f}, \bm{3'}}^\phi + W_{\mathrm{f}, \xi} +
 W_{\mathrm{f}, \bm{3'}}^\psi + W_{\mathrm{f}, \xi'},\label{eq_super_all}\\
 W_{\mathrm{f}, \ell} &= \frac{1}{\Lambda^3} \left[
 h_1 \left( \phi_{\ell}^3 \right)_{\bm{1'}}^2
 + h_2 \left( \phi_{\ell}^2 \right)_{\bm{1}}
 \left( \left( \phi_{\ell}^2 \right)_{\bm{3'}}^2 \right)_{\bm{1}}
 \right],  \label{eq_super_ell}
 \\
 W_{\mathrm{f}, \bm{1}} &= \frac{g_0}{\Lambda^3} \phi_{\bm{1}}^6
 ~~;~~
 W_{\mathrm{f}, \xi} = \frac{g_\xi}{\Lambda^3} \xi_{\bm{1}}^6
 ~~;~~
 W_{\mathrm{f}, \xi'} = \frac{g_{\xi'}}{\Lambda^3} \xi_{\bm{1}}^{'6},\\
 W_{\mathrm{f}, \bm{2}} &= \frac{1}{\Lambda^3} \left[
 g_1 \left( \phi_{\bm{2}}^2 \right)_{\bm{1}}^3
 + g_2 \left( \phi_{\bm{2}}^3 \right)_{\bm{1}}^2
 \right],\\
 W_{\mathrm{f}, \bm{3'}}^\phi &= \frac{1}{\Lambda^3} \left[
 g_3 \left( \phi_{\bm{3'}}^2 \right)_{\bm{1}}^3
 + g_4 \left( \left[ \phi_{\bm{3'}}^5 \right] \phi_{\bm{3'}} \right)_{\bm{1}}
 \right],\\
 W_{\mathrm{f}, \bm{3'}}^\psi &= \frac{1}{\Lambda^3} \left[
 g_5 \left( \psi_{\bm{3'}}^2 \right)_{\bm{1}}^3
 + g_6 \left( \left[ \psi_{\bm{3'}}^5 \right] \psi_{\bm{3'}} \right)_{\bm{1}}
 \right],
 \label{eq_super}
\end{align}
where the block parenthesis for $\phi_{\bm{3'}}$ is recursively defined as
\begin{align}
 [\phi_{\bm{3'}}] &\equiv \phi_{\bm{3'}},\\
 [\phi_{\bm{3'}}^n] &\equiv \left( [\phi_{\bm{3'}}^{n-1}] \phi_{\bm{3'}} \right)_{\bm{3'}},
\end{align}
and the same applies to $\psi_{\bm{3'}}$. 
The above definition uniquely specifies the way of contraction in each term. 
Some comments are in order.  First, we assume that the superpotential of each flavon is separated as 
$W_{\mathrm{f},\phi} = W_{\mathrm{f},\bm{1}} + W_{\mathrm{f},\bm{2}} + W_{\mathrm{f},\bm{3'}}^\phi$.  
Although the configuration \eqref{eq_vev_nu} remains an extremum even if we allow to include mixings,
as shown in the previous section, we focus on the case without such mixings for simplicity.  
Second, many other ways of contraction of the flavon with the same power are also allowed, 
but it may be natural that some of these possible terms have relatively large numerical
coefficients. The superpotential presented in (\ref{eq_super_ell})--(\ref{eq_super}) should be 
regarded as one of the examples.  Our purpose here is to give an existence proof of
consistent parameter regions where flavons are correctly stabilized.

In order to stabilize flavon fields with non-zero VEVs, we add negative SUSY breaking masses for flavon fields
\begin{align}
 V_{\mathrm{SB}} =
&-m_{\ell}^2 \sum_{i=1}^{3} \left|\phi_{\ell,i}\right|^2
 -m_{\bm{1}}^2 \left|\phi_{\bm{1}}\right|^2
 -m_{\bm{2}}^2 \sum_{i=1}^{2} \left|\phi_{\bm{2},i}\right|^2
 -m_{\bm{3'}}^2 \sum_{i=1}^{3} \left|\phi_{\bm{3'},i}\right|^2
 -m_{\xi}^2 \left| \xi_{\bm{1}} \right|^2 \notag \\
&-m_{\psi}^2 \sum_{i=1}^{3} \left|\psi_{\bm{3'},i}\right|^2
 -m_{\xi'}^2 \left| \xi'_{\bm{1}} \right|^2.
\end{align}
Then, the full flavon scalar potential is given by
\begin{align}
 V = 
&\sum_{i=1}^{3} \left| \frac{\partial W_{\mathrm{f}, \ell}}{\partial \phi_{\ell,i}} \right|^2
 + \left| \frac{\partial W_{\mathrm{f}, \bm{1}}}{\partial \phi_{\bm{1}}} \right|^2
 + \sum_{i=1}^{2} \left| \frac{\partial W_{\mathrm{f}, \bm{2}}}{\partial \phi_{\bm{2},i}} \right|^2
 + \sum_{i=1}^{3} \left| \frac{\partial W_{\mathrm{f}, \bm{3'}}^\phi}{\partial \phi_{\bm{3'},i}} \right|^2
 + \left| \frac{\partial W_{\mathrm{f}, \xi}}{\partial \xi_{\bm{1}}} \right|^2 \notag \\
&+ \sum_{i=1}^{3} \left| \frac{\partial W_{\mathrm{f}, \bm{3'}}^\psi}{\partial \psi_{\bm{3'},i}} \right|^2
 + \left| \frac{\partial W_{\mathrm{f}, \xi'}}{\partial \xi'_{\bm{1}}} \right|^2
 + V_{\mathrm{SB}}
 + V_A,
\end{align}
where $V_A$ is the so-called $A$-term potential induced by the supergravity effects
\begin{align}
 V_A = \frac{3A}{\Lambda^3} &\left[
 h_1 \left( \phi_{\ell}^3 \right)_{\bm{1'}}^2
 + h_2 \left( \phi_{\ell}^2 \right)_{\bm{1}}
 \left( \left( \phi_{\ell}^2 \right)_{\bm{3'}}^2 \right)_{\bm{1}}
 \right.
 \notag \\
 &+
 g_0 \phi_{\bm{1}}^6
 + g_1 \left( \phi_{\bm{2}}^2 \right)_{\bm{1}}^3
 + g_2 \left( \phi_{\bm{2}}^3 \right)_{\bm{1}}^2
 + g_3 \left( \phi_{\bm{3'}}^2 \right)_{\bm{1}}^3
 + g_4 \left( \left[ \phi_{\bm{3'}}^5 \right] \phi_{\bm{3'}} \right)_{\bm{1}}
 + g_\xi  \xi_{\bm{1}}^6
 \notag \\
 &+\left.
 g_5 \left( \psi_{\bm{3'}}^2 \right)_{\bm{1}}^3
 + g_6 \left( \left[ \psi_{\bm{3'}}^5 \right] \psi_{\bm{3'}} \right)_{\bm{1}}
 + g_{\xi'} \xi_{\bm{1}}^{'6}
 \right] + \mathrm{h.c.}
\end{align}
where $|A| \sim m_{3/2}$ with $m_{3/2}$ being the gravitino mass.  Now we will check that this 
model possesses a desired set of vacuum expectation values given in \eqref{eq_vev_ell} and 
\eqref{eq_vev_nu} depending on the choice of parameters $h_i~(i=1,2)$ and
$g_i~(i=0, ..., 6)$.  For simplicity, we take all coupling constants $h_i$ and $g_i$ real.  
We will discuss below the stabilization of $\phi_{\bf 1}$, $\phi_{\ell}$, $\phi_{\bm{2}}$,
$\phi_{\bm{3'}}$ and $\psi_{\bm{3'}}$ in order.

\subsection{Potential of $\phi_{\bf 1}$}

First, we consider the stabilization of $\phi_{\bf 1}$. For the moment, we neglect the contribution from 
$V_A$, which is justified if $m_{3/2} \ll m_{\bf 1}$. We find a minimum of the potential at
\begin{align}
	v_{\bf 1} =  \left( \frac{1}{180} \right)^{1/8} \left( \frac{m_{\bf 1} \Lambda^3}{|g_0|} \right)^{1/4}.  \label{v1}
\end{align}
Expanding $\phi_{\bf 1}$ around the VEV as
\begin{align}
	\phi_{\bf 1} = v_1+ \frac{1}{\sqrt{2}} \left( \phi_{\bf 1}^R +i \phi_{\bf 1}^I \right),
\end{align}
we find that $ \phi_{\bf 1}^R$ has a mass of $2\sqrt{2} m_{\bf 1}$. On the other hand, $\phi_{\bf 1}^I$ is 
massless as far as the $A$-term contribution is neglected. This is expected since the scalar potential has 
a global U(1) symmetry and there should be a Nambu-Goldstone mode when we neglect the $A$-term. 
After including the $A$-term, $\phi_{\bf 1}^I$ obtains positive mass squared of the order of $|A m_{\bf 1}|$
for $A>0$ $(A<0)$ when $g_0<0$ $(g_0>0)$. Notice that the stabilizations of $\xi_{\bf 1}$ and $\xi'_{\bf 1}$ 
are completely parallel to $\phi_{\bf 1}$.

\subsection{Potential of $\phi_{\ell}$}  \label{sec:phiell}

\begin{figure}
 \begin{subfigure}{0.495\linewidth}
  \includegraphics[width=\linewidth]{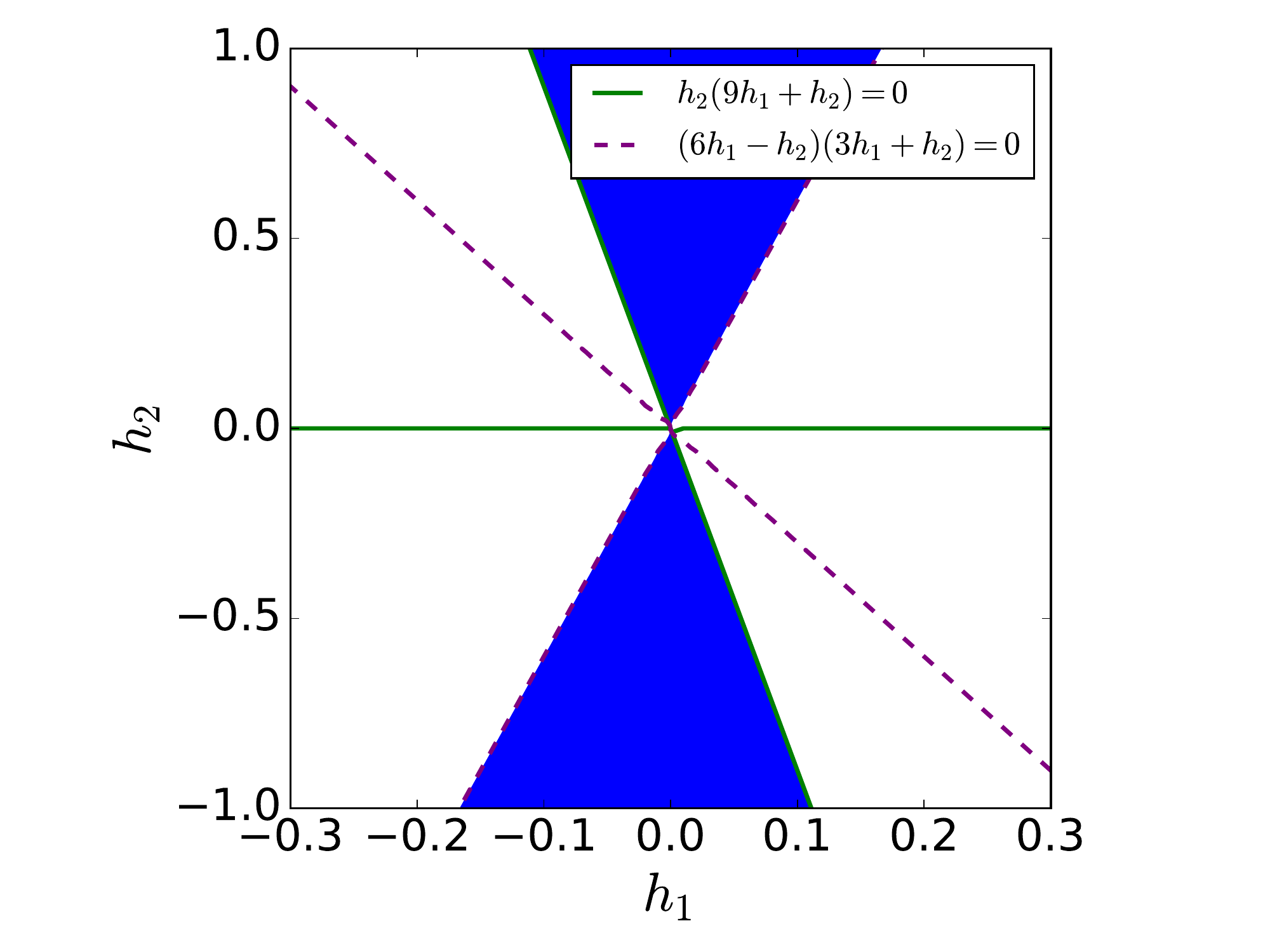}\\
  \includegraphics[width=\linewidth]{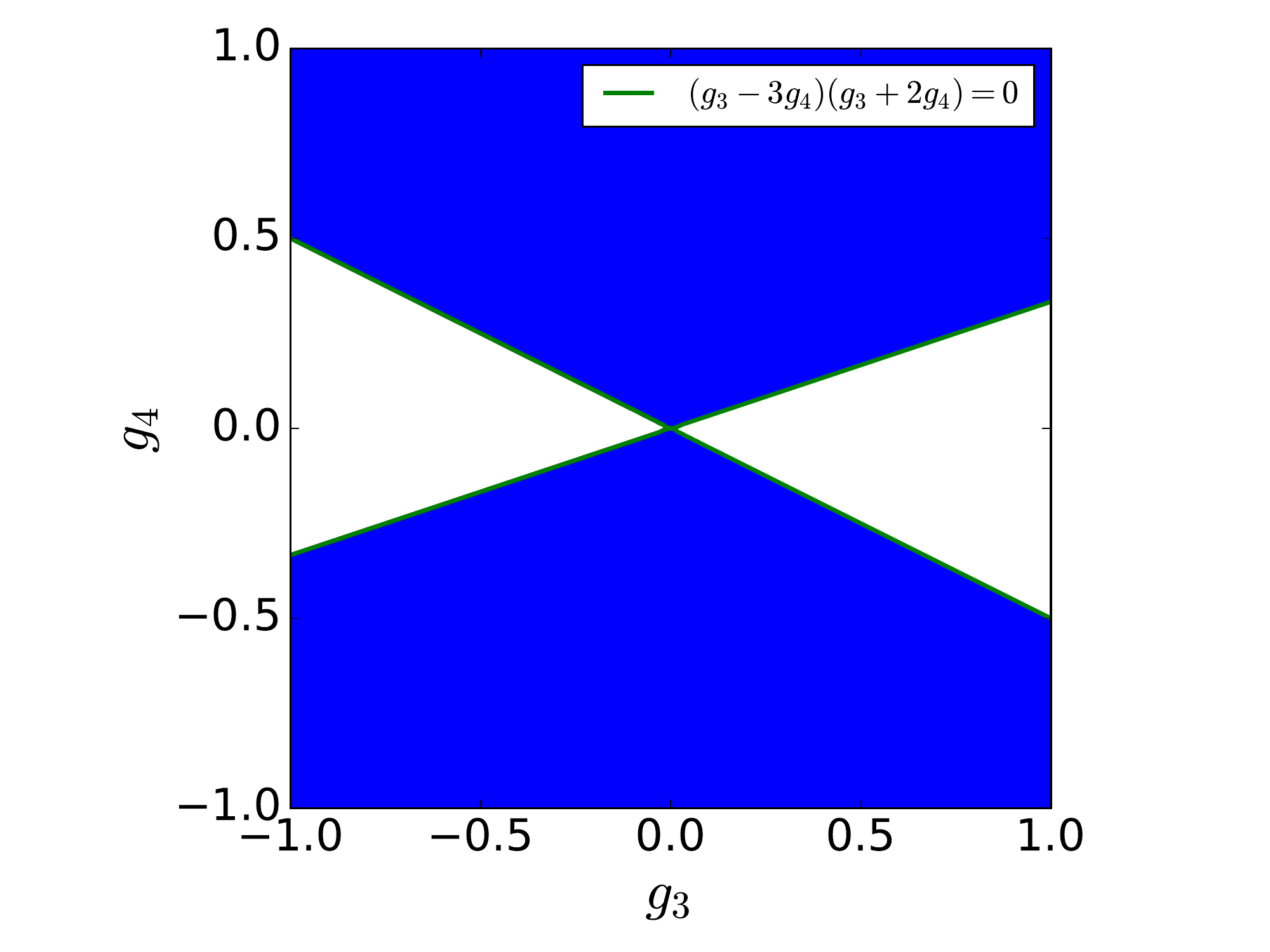}
 \end{subfigure}
 \begin{subfigure}{0.495\linewidth}
  \includegraphics[width=\linewidth]{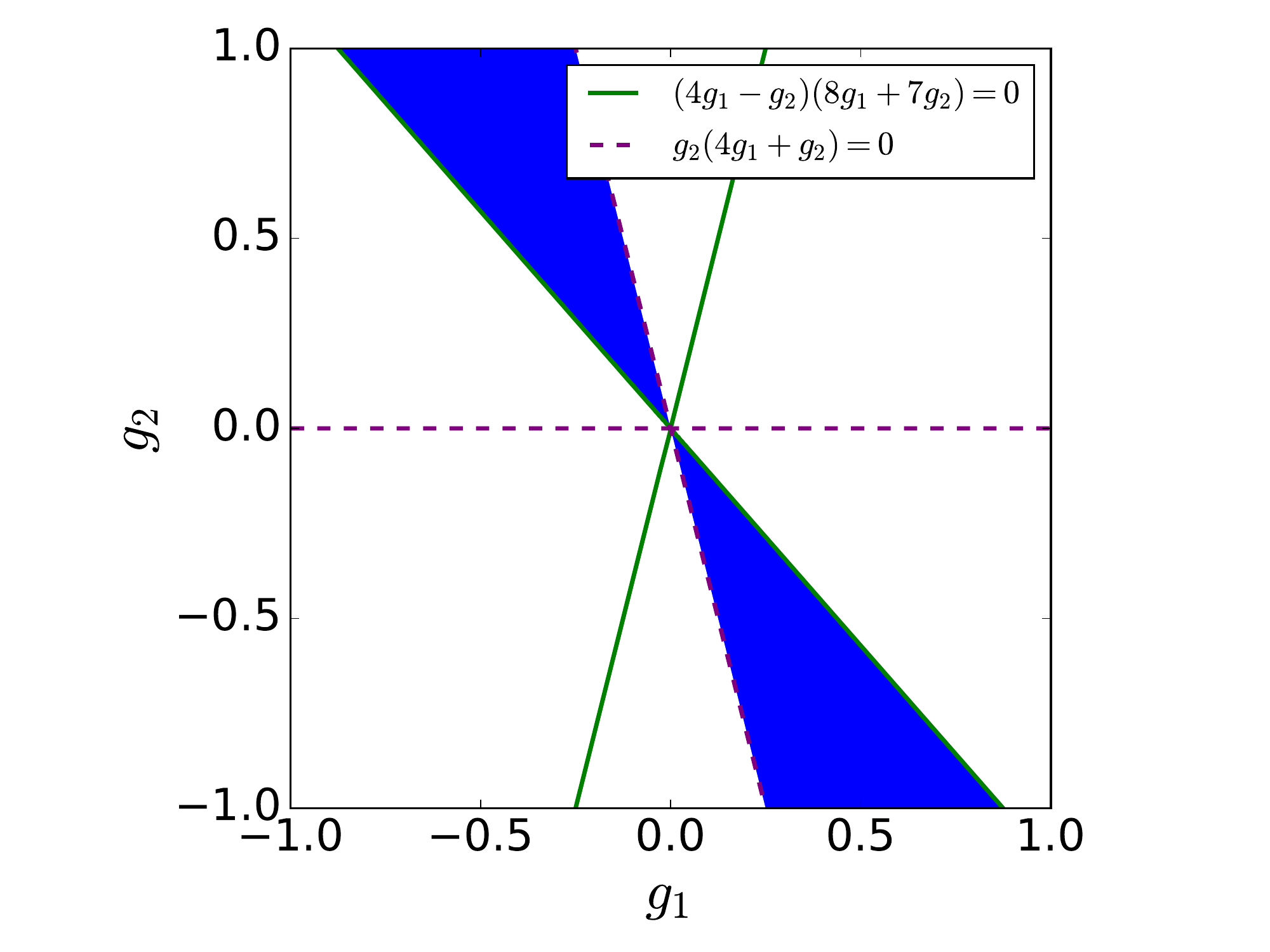}\\
  \includegraphics[width=\linewidth]{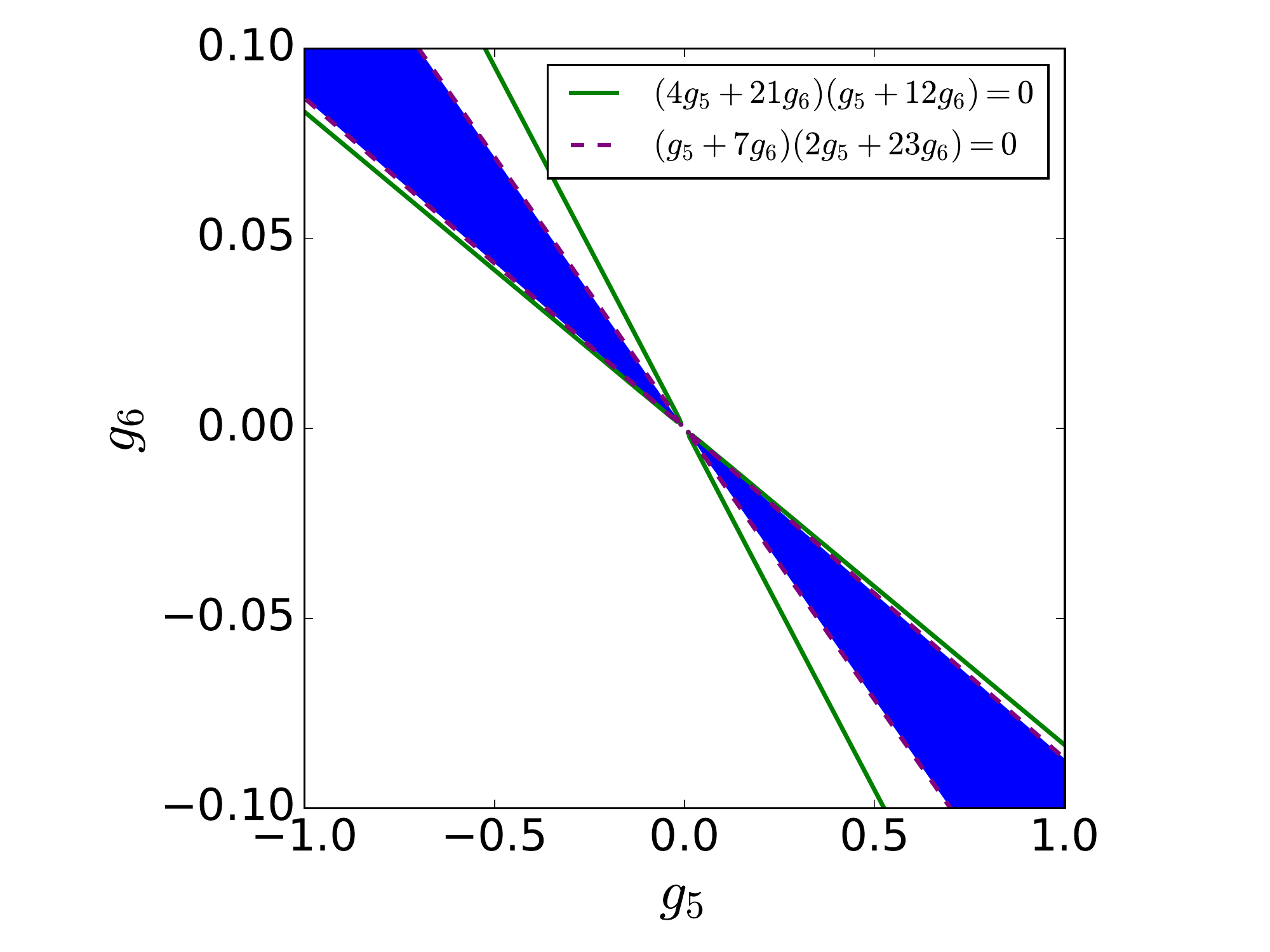}
 \end{subfigure}
 \caption{The stabilization conditions of the flavon fields imposed on the parameter space. 
 In each figure, the blue region is allowed by the conditions.  Figures correspond to the parameter 
 space of the $\phi_\ell$ sector (upper left), the $\phi_{\bm{2}}$ sector (upper right), the 
 $\phi_{\bm{3'}}$ sector (lower left), and the $\psi_{\bm{3'}}$ sector (lower right).  For the relevant 
 conditions, see \eqref{eq_stab_ell}, \eqref{eq_stab_2}, \eqref{eq_stab_3p}, \eqref{eq_stab_psi_1}, 
 and \eqref{eq_stab_psi_2}.}
 \label{fig_stab_cond}
\end{figure}

Here, we consider the stabilization of $\phi_{\ell}$. As shown in Sec.~\ref{sec:general}, we can
always find an extremum of the form of \eqref{eq_vev_ell} with
\begin{align}
 v_\ell = \left( \frac{1}{2880} \right)^{1/8} \left( \frac{m_\ell \Lambda^3}{|h_1|} \right)^{1/4}.  \label{vell}
\end{align}
Actually, the phase of $v_\ell$ is fixed after the $A$-term is taken into account, as we shall see below.
If all the mass eigenvalues of the flavon fluctuations around this extremum are positive, this extremum is, 
in fact, a (local) minimum. In order to see this, we expand the flavon fields as
\begin{align}
 \phi_{\ell,1} = \frac{1}{\sqrt{2}} \left( \phi_{\ell,1}^R +i \phi_{\ell,1}^I \right),
 ~~
 \phi_{\ell,2} = v_\ell + \frac{1}{\sqrt{2}} \left( \phi_{\ell,2}^R +i \phi_{\ell,2}^I \right),
 ~~
 \phi_{\ell,3} = \frac{1}{\sqrt{2}} \left( \phi_{\ell,3}^R +i \phi_{\ell,3}^I \right),
\end{align}
Then we find out that the mass matrix becomes block diagonal, where the mass for the real and imaginary
parts of $\phi_{\ell}$ are separated with each other.  As for the real part of the flavon fields, it can be seen 
that $\phi_{\ell,2}^R$ is a mass eigenstate with mass $2\sqrt{2}m_\ell$, while the other two modes
possess a mass matrix of the form of \eqref{eq_m2sq} with
\begin{align}
 B &= -\frac{4(9h_1^2-3h_1h_2-h_2^2)}{45h_1^2} m_\ell^2,\\
 C &= -\frac{4(3h_1+2h_2)}{15h_1} m_\ell^2.
\end{align}
Then, the required condition for both of the two mass eigenvalues to be positive is
\begin{align}
 h_2(9h_1+h_2) > 0
 ~~\mathrm{and}~~
 (6h_1-h_2)(3h_1+h_2) < 0.\label{eq_stab_ell}
\end{align}
Next, let us turn to the imaginary part of the flavon fields $\phi_{\ell,i}^I (i=1,3)$. They possess 
the mass matrix that is the same as that of the real part but the sign of $C$ is flipped.  As a
result, the same condition \eqref{eq_stab_ell} ensures the stability of the corresponding mass 
eigenstates.  On the other hand, $\phi_{\ell,2}^I$ becomes massless at this approximation 
but $V_A$ gives the leading contribution to its mass. If we write the VEV of the flavon as 
$\left<\phi_{\ell,2}\right> = v_\ell e^{i\varphi}$, the potential is minimized at 
$\varphi = 2n\pi/3$ $(n=0,1,2)$ for $A>0$ ($A<0$) when $h_1<0$ ($h_1>0$). In Eq.~(\ref{vell}) 
we just chose $n=0$ solution.  A similar argument should be understood in the following 
analysis of $\phi_{\bf 2}, \phi_{\bf 3'}$ and $\psi_{\bf 3'}$.
In the upper left part of Fig.~\ref{fig_stab_cond}, we show the allowed region in the
$h_1$ vs. $h_2$ plane as a blue region.

\subsection{Potential of $\phi_{\textbf{2}}$}  \label{sec:phi2}

Now, we consider the $\phi_{\bm{2}}$ sector of the potential. There exists an extremum of the form 
of \eqref{eq_vev_nu}, irrespective of the choice of parameters, with
\begin{align}
 v_2 = \left( \frac{1}{720} \right)^{1/8}
 \left( \frac{m_{\bm{2}} \Lambda^3}{|2g_1+g_2|} \right)^{1/4}.
\end{align}
Expanding the flavon fields as
\begin{align}
 \phi_{\bm{2},1} = v_2 + \frac{1}{\sqrt{2}}
 \left( \phi_{\bm{2},1}^R + i \phi_{\bm{2},1}^I \right),
 ~~
 \phi_{\bm{2},2} = v_2 + \frac{1}{\sqrt{2}}
 \left( \phi_{\bm{2},2}^R + i \phi_{\bm{2},2}^I \right),
\end{align}
we find the mass matrix being block diagonal when we neglect the contribution from $V_A$.  
The mass matrix for the real part is given by \eqref{eq_m2sq} with
\begin{align}
 B &= \frac{128g_1^2 + 140g_1g_2 + 47g_2^2}{10(2g_1+g_2)^2} m_{\bm{2}}^2,\\
 C &= \frac{3(64g_1^2 + 60g_1g_2 + 11g_2^2)}{10(2g_1+g_2)^2} m_{\bm{2}}^2,
\end{align}  
and the condition for both mass eigenvalues to be positive is given by
\begin{align}
 (4g_1-g_2) (8g_1+7g_2) < 0. \label{eq_stab_2}
\end{align}
On the other hand, the mass matrix for the imaginary part is also given by \eqref{eq_m2sq} but with
\begin{align}
 B = -\frac{9g_2(4g_1+g_2)}{10(2g_1+g_2)^2} m_{\bm{2}}^2 = -C.
\end{align}
As mentioned in Appendix~\ref{sec_diag}, this mass matrix possesses a massless mode and the 
condition for another mass eigenstate to have positive mass squared is
\begin{align}
 g_2 (4g_1+g_2) < 0.
\end{align}
As for the massless mode, the potential $V_A$ neglected so far gives it a non-zero mass-squared, 
whose sign becomes positive if $A>0$ ($A<0$) when $2g_1+g_2<0$ ($2g_1+g_2>0$). In the upper 
right part of Fig.~\ref{fig_stab_cond}, we show the allowed region in the $g_1$ vs. $g_2$ plane.

\subsection{Potential of $\phi_{\textbf{3}'}$}  \label{sec:phi3}

Let us move on to the $\phi_{\bm{3'}}$ sector of the potential.  We can find an extremum at the position 
of \eqref{eq_vev_nu}, with $v_{3'}$ given by
\begin{align}
 v_{3'} = \left( \frac{1}{14580} \right)^{1/8}
 \left( \frac{m_{\bm{3'}} \Lambda^3}{|g_3|} \right)^{1/4}.
\end{align}
Expanding the flavon fields around the extremum as
\begin{align}
 \phi_{\bm{3'},1} = v_{3'} + \frac{1}{\sqrt{2}} ( \phi_{\bm{3'},1}^R + i\phi_{\bm{3'},1}^I ),
 ~~
 \phi_{\bm{3'},2} = v_{3'} + \frac{1}{\sqrt{2}} ( \phi_{\bm{3'},2}^R + i\phi_{\bm{3'},2}^I ),
 ~~
 \phi_{\bm{3'},3} = v_{3'} + \frac{1}{\sqrt{2}} ( \phi_{\bm{3'},3}^R + i\phi_{\bm{3'},3}^I ),
\end{align}
we obtain a block diagonal mass matrix for flavon fields when we neglect the small contribution 
from $V_A$.  The mass matrix for the real part of $\phi_{\bm{3'}}$ takes the form of \eqref{eq_m3sq} with
\begin{align}
 D &= \frac{8(5g_3^2+18g_3g_4+36g_4^2)}{15g_3^2} m_{\bm{3'}}^2,\\
 E &= \frac{8(g_3-3g_4)(5g_3+6g_4)}{15g_3^2} m_{\bm{3'}}^2,\\
 F &= \frac{4(7g_3^2+72g_4^2)}{15g_3^2} m_{\bm{3'}}^2,\\
 G &= \frac{4(13g_3^2+18g_3g_4-36g_4^2)}{15g_3^2} m_{\bm{3'}}^2.
\end{align}
We find that, if the coupling parameters satisfy the conditions
\begin{align}
(g_3-3g_4)(g_3+2g_4) < 0,
 \label{eq_stab_3p}
\end{align}
all the eigenvalues become positive.  For the imaginary part of $\phi_{\bm{3'}}$, the same conditions 
ensure the stability of all the eigenstates but one that becomes massless.  Again, the dominant
contribution to its mass comes from $V_A$, whose sign becomes positive if $A>0$ ($A<0$) when 
$g_3<0$ ($g_3>0$). In the lower left part of Fig.~\ref{fig_stab_cond}, we show the allowed region 
in the $g_3$ vs. $g_4$ plane.

\subsection{Potential of $\psi_{\textbf{3}'}$}  \label{sec:psi3}

The form of the potential of $\psi_{\bm{3'}}$ is exactly the same as that of $\phi_{\bm{3'}}$.  However, 
as we will see below, it is possible for the potential to possess a required form of the vacuum depending 
on the choice of the coupling parameters.  As in the other sectors, we can find an extremum of the form 
of \eqref{eq_vev_nu} irrespective of the coupling parameters, with $v_\psi$ expressed as
\begin{align}
 v_\psi = \left( \frac{1}{2880} \right)^{1/8}
 \left( \frac{m_\psi \Lambda^3}{|g_5+9g_6|} \right)^{1/4}.
\end{align}
We expand the flavon fields around the extremum as
\begin{align}
 \psi_{\bm{3'},1} = \frac{1}{\sqrt{2}} ( \psi_{\bm{3'},1}^R + i\psi_{\bm{3'},1}^I ),
 ~~
 \psi_{\bm{3'},2} = v_\psi + \frac{1}{\sqrt{2}} ( \psi_{\bm{3'},2}^R + i\psi_{\bm{3'},2}^I ),
 ~~
 \psi_{\bm{3'},3} = -v_\psi + \frac{1}{\sqrt{2}} ( \psi_{\bm{3'},3}^R + i\psi_{\bm{3'},3}^I ),
\end{align}
and calculate the mass matrix of the real and imaginary parts of $\psi_{\bm{3'}}$.  For the real part, 
we obtain a matrix of the form of \eqref{eq_m3sq} with
\begin{align}
 D &= \frac{-(8g_5^2+138g_5g_6+423g_6^2)}{5(g_5+9g_6)^2} m_\psi^2,\\
 E &= \frac{-9g_6(2g_5+15g_6)}{5(g_5+9g_6)^2} m_\psi^2,\\
 F &= \frac{16g_5^2+282g_5g_6+1341g_6^2}{5(g_5+9g_6)^2} m_\psi^2,\\
 G &= \frac{-3(8g_5^2+146g_5g_6+633g_6^2)}{5(g_5+9g_6)^2} m_\psi^2,
\end{align}
all of whose eigenvalues become positive if
\begin{align}
 (4g_5+21g_6)(g_5+12g_6) < 0.
 \label{eq_stab_psi_1}
\end{align}
Similarly, the mass matrix for the imaginary part is given by
\begin{align}
 D &= \frac{-9g_6(2g_5-g_6)}{5(g_5+9g_6)^2} m_\psi^2,\\
 E &= \frac{27g_6(2g_5+19g_6)}{5(g_5+9g_6)^2} m_\psi^2,\\
 F &= G = \frac{9g_6(2g_5+29g_6)}{5(g_5+9g_6)^2} m_\psi^2,
\end{align}
and the conditions for all the eigenvalues to be positive are
\begin{align}
 (g_5+7g_6)(2g_5+23g_6) < 0,
 \label{eq_stab_psi_2}
\end{align}
together with the condition for the stabilization of one of the eigenstates by $V_A$: $A>0$ ($A<0$) 
when $g_5+9g_6>0$ ($g_5+9g_6<0$). In the lower right part of Fig.~\ref{fig_stab_cond}, we show 
the allowed region in the $g_5$ vs. $g_6$ plane.

\subsection{Fermion sector}

Finally, we briefly comment on the masses of fermionic partners of the flavon fields (flavinos). 
By diagonalizing their mass matrices, we can easily check that all the mass eigenstates generally 
possess masses of the order of the SUSY breaking mass scale.  Below we list all the mass 
eigenstates and the corresponding mass eigenvalues. For the $S_4$ singlet flavon sector, we have
\begin{align}
	&\widetilde\phi_{\bm{1}} ~~;~~m = \mathrm{sgn}(g_0)\sqrt{5} m_{\bm{1}},\\
	&\widetilde\xi_{\bm{1}} ~~;~~m = \mathrm{sgn}(g_\xi)\sqrt{5} m_{\bm{\xi}},\\
	&\widetilde\xi'_{\bm{1}} ~~;~~m = \mathrm{sgn}(g_{\xi'})\sqrt{5} m_{\bm{\xi'}},
\end{align}
where the tilde denotes the fermionic superpartner of any scalar field. 
For the $\phi_\ell$ sector, we have
\begin{align}
 \widetilde{\phi}_{\ell,2}~~&;~~m = \mathrm{sgn}(h_1) \sqrt{5} m_\ell,\\
 \frac{1}{\sqrt{2}} \left( \widetilde{\phi}_{\ell,1} \pm \widetilde{\phi}_{\ell,3} \right)
 ~~&;~~m = \pm \frac{(3h_1+2h_2)}{3\sqrt{5} |h_1|} m_\ell.
\end{align}
For the $\phi_{\bm{2}}$ sector, we have
\begin{align}
 \frac{1}{\sqrt{2}} \left( \widetilde{\phi}_{\bm{2},1} + \widetilde{\phi}_{\bm{2},2} \right)
 ~~&;~~m = \mathrm{sgn} (2g_1 + g_2) \sqrt{5} m_{\bm{2}},\\
 \frac{1}{\sqrt{2}} \left( \widetilde{\phi}_{\bm{2},1} - \widetilde{\phi}_{\bm{2},2} \right)
 ~~&;~~m = \frac{2(g_1 - g_2)}{\sqrt{5} |2g_1 + g_2|} m_{\bm{2}}.
\end{align}
Finally, for the $\phi_{\bm{3'}}$ sector, we obtain
\begin{align}
 \frac{1}{\sqrt{3}} \left( \widetilde{\phi}_{\bm{3'},1} + \widetilde{\phi}_{\bm{3'},2} + \widetilde{\phi}_{\bm{3'},3} \right)
 ~~&;~~m = \sqrt{5} m_{\bm{3'}},\\
 \frac{1}{\sqrt{2}} \left( \widetilde{\phi}_{\bm{3'},2} - \widetilde{\phi}_{\bm{3'},3} \right)
 ~~&;~~m = - \frac{(g_3 + 12g_4)}{\sqrt{5} |g_3|} m_{\bm{3'}},\\
 \frac{1}{\sqrt{6}} \left( -2\widetilde{\phi}_{\bm{3'},1} + \widetilde{\phi}_{\bm{3'},2} + \widetilde{\phi}_{\bm{3'},3} \right)
 ~~&;~~m = \frac{(g_3 + 12g_4)}{\sqrt{5} |g_3|} m_{\bm{3'}},
\end{align}
and for $\psi_{\bm{3'}}$ sectors,
\begin{align}
 \frac{1}{\sqrt{3}} \left( \widetilde{\psi}_{\bm{3'},1} + \widetilde{\psi}_{\bm{3'},2} + \widetilde{\psi}_{\bm{3'},3} \right)
 ~~&;~~m = \frac{(g_5 + 24g_6)}{\sqrt{5} |g_5 + 9g_6|} m_\psi,\\
 \frac{1}{\sqrt{2}} \left( \widetilde{\psi}_{\bm{3'},2} - \widetilde{\psi}_{\bm{3'},3} \right)
 ~~&;~~m = -\mathrm{sgn} (g_5 + 24g_6) \sqrt{5} m_\psi,\\
 \frac{1}{\sqrt{6}} \left( -2\widetilde{\psi}_{\bm{3'},1} + \widetilde{\psi}_{\bm{3'},2} + \widetilde{\psi}_{\bm{3'},3} \right)
 ~~&;~~m = \frac{(g_5 - 3g_6)}{\sqrt{5} |g_5 + 9g_6|} m_\psi.
\end{align}
All flavinos have masses of the order of the SUSY breaking mass and hence it may be possible that one 
of them is the lightest SUSY particle and a candidate of dark matter. There are several ways to produce 
flavinos in the early universe, for example, by thermal scattering of minimal SUSY standard model (MSSM) 
particles or the decay of MSSM particles, gravitino, flavons and so on.

\section{Conclusions and discussion} \label{sec:conc}

An $S_4$ flavor model can lead to the so-called TM$_1$ pattern of
neutrino mixings consistent with current experimental data if all the
flavons are stabilized appropriately. We have explicitly constructed a
model in which all flavons have VEVs with the desired alignment
structure in a simple way.  The flavon stabilization is achieved by the
balance between the tachyonic SUSY breaking mass and the higher
dimensional terms in the potential. In our model, we do not need any
additional field (such as the driving fields) in order to stabilize
flavons. In this sense, our model is very simple.  In addition, although
we study an $S_4$ model in this paper, this mechanism is universal and
can be applied to many flavor models based on discrete flavor symmetry.

Having seen that the desired flavon VEV alignments \eqref{eq_vev_ell} and \eqref{eq_vev_nu} can be 
obtained in our setup, we shortly discuss the implications for the cosmological domain wall problem.
During inflation, the flavons are stabilized due to the negative Hubble-induced mass terms~\cite{Dine:1995uk} 
instead of the tachyonic SUSY breaking mass terms, but the VEV alignment structure is exactly the same 
as that by the latter. Once the flavons settle down in the desired minimum of the potential during inflation, 
they remain trapped in the varying minimum after inflation when the Hubble parameter $H$ gradually decreases. 
The flavons finally get into the present vacuum when $H$ becomes comparable 
to the SUSY breaking mass. One can show that flavons do not overshoot the origin of the field space 
during this whole cosmological dynamics as shown in Ref.~\cite{Chigusa:2018yua}. Thus in this scenario, 
the discrete flavor symmetry is already spontaneously broken during inflation and never restored thereafter, 
which implies that there is no cosmological domain wall problem.

So far we have neglected thermal effects on the flavon potential. The existence of high-temperature 
plasma in the early universe can affect the flavon potential that might lead to the symmetry restoration. 
For concreteness, let us consider a standard cosmological scenario that the universe enters in the matter 
domination era due to the coherent inflaton oscillation after the end of inflation and finally the inflaton 
decays and the radiation-dominated universe begins at $T=T_{\rm R}$, where $T$ is the radiation 
temperature and $T_{\rm R}$ is the reheating temperature. In our model, flavons do not have 
renormalizable interactions with MSSM fields. Still, non-renormalizable interactions may give rise 
to sizable thermal effects. The dominant thermal effect on the flavon potential comes from the tau 
Yukawa coupling in (\ref{W_yukawa}). It arises at the two-loop order,
\begin{align}
	V_T \sim \frac{y_\tau^2 T^4}{\Lambda^2}|\phi_\ell|^2.
\end{align}
In order for this thermal potential not to affect the flavon dynamics significantly, we demand that the thermal 
potential is always subdominant compared with either the Hubble mass term $H^2|\phi_\ell|^2$ or the SUSY 
breaking mass term $m_\ell^2|\phi_\ell|^2$. Assuming $T_{\rm R} \lesssim \sqrt{m_\ell M_{\rm P}}$, where 
$M_{\rm P}$ is the reduced Planck scale, it is sufficient to demand that 
\begin{align}
	\left.\frac{y_\tau^2 T^4}{\Lambda^2}\right|_{H=m_\ell}\lesssim m_\ell^2 ~~~\leftrightarrow ~~~
	y_\tau^2 \left( \frac{T_{\rm R}}{10^6\,{\rm GeV}} \right)^2 \left( \frac{10^6\,{\rm GeV}}{m_\ell} \right)
	\left( \frac{10^{12}\,{\rm GeV}}{\Lambda} \right)^2 \lesssim 1,
\end{align}
where we have used $T^4 \sim T_{\rm R}^2 H M_{\rm P}$ before the completion of the reheating. 
If this condition is satisfied, thermal effects on the flavon dynamics are safely neglected.\footnote{
Although it does not lead to the symmetry restoration, a small amount of flavon oscillation around its potential 
minimum may be induced~\cite{Buchmuller:2004xr,Nakayama:2008ks,Lillard:2018zts,Hagihara:2018uix}.
} Thermal effects on the other flavons, $\phi_{\bf 1}$, $\phi_{\bf 2}$, $\phi_{\bf 3'}$, and $\psi_{\bf 3'}$, are suppressed 
by an additional power of $(T/\Lambda)^2$ and hence are negligible.  Therefore, as far as the reheating 
temperature is not too high, our model is cosmologically viable.

\section*{Acknowledgments}

This work was supported by the JSPS KAKENHI Grant (No. 17J00813 [SC]), Grant-in-Aid for Scientific 
Research C (No.18K03609 [KN]) and Innovative Areas (No.15H05888 [KN], No.17H06359 [KN]).

\appendix

\section{Notes on $S_4$ representations} \label{app:S4}

In this Appendix, we summarize the representation of the $S_4$ group and product rules.
Our convention is the same as e.g. Appendix of Ref.~\cite{Ding:2013hpa}.

All the elements of the $S_4$ group can be written as a product of three elements often called 
$S$, $U$, and $T$, which generates $Z_2$, another $Z_2$, and $Z_3$ subgroup of $S_4$, respectively. 
$S_4$ has five different representations: $\bm{1}$, $\bm{1'}$, $\bm{2}$, $\bm{3}$, and $\bm{3'}$, where 
the number denotes the dimension of each representation.  In doublet representation $\bm{2}$,
representation matrices for generators are given as
\begin{align}
 S=\begin{pmatrix}
    1 & 0 \\
    0 & 1
   \end{pmatrix},~~~
 U=\begin{pmatrix}
    0 & 1 \\
    1 & 0
   \end{pmatrix},~~~
 T=\begin{pmatrix}
    \omega & 0 \\
    0 & \omega^2
   \end{pmatrix},
   \label{SUT2}
\end{align}
where we define $\omega \equiv e^{2 \pi i / 3}$.  In triplet representations $\bm{3}$ and 
$\bm{3'}$, corresponding matrices are
\begin{align}
 S=\frac{1}{3}\begin{pmatrix}
	       -1 & 2 & 2\\
	       2 & -1 & 2\\
	       2 & 2 & -1
	      \end{pmatrix},~~~
 U=\mp\begin{pmatrix}
       1 & 0 & 0\\
       0 & 0 & 1\\
       0 & 1 & 0
      \end{pmatrix},~~~
 T=\begin{pmatrix}
    1 & 0 & 0\\
    0 & \omega^2 & 0\\
    0 & 0 & \omega
   \end{pmatrix},
   \label{SUT3}
\end{align}
where $-$ ($+$) for the $\bm{3}$ ($\bm{3'}$) representation.

In order to show the product rules in this basis, we define a non-trivial singlet $p'$ in $\bm{1'}$, 
a doublet $a=(a_1, a_2)$, and two triplets $b=(b_1,b_2,b_3)$ in $\bm{3}$ and $b'=(b_1',b_2',b_3')$
in $\bm{3'}$.  We also use a tilde in order to use another multiplet in the same representation. 
Firstly, products with $\bm{1'}$ are decomposed as $\bm{1'} \times \bm{1'} = \bm{1}$, 
$\bm{2} \times \bm{1'} = \bm{2}$, $\bm{3} \times \bm{1'} = \bm{3'}$, and 
$\bm{3'} \times \bm{1'} = \bm{3}$: in components,
\begin{align}
 (p' \tilde{p}')_{\bm{1}} = p' \tilde{p}',~~~
 (p' a)_{\bm{2}} = \begin{pmatrix}
	      p' a_1 \\
	      - p' a_2
	     \end{pmatrix},~~~
 (p' b)_{\bm{3'}} = \begin{pmatrix}
	       p' b_1 \\
	       p' b_2 \\
	       p' b_3
	      \end{pmatrix},~~~
 (p' b')_{\bm{3}} = \begin{pmatrix}
	      p' b_1' \\
	      p' b_2' \\
	      p' b_3'
	     \end{pmatrix},
\end{align}
where the parenthesis denotes the contraction of several representations that as a whole transforms 
as a representation denoted by the subscript.  Next, the product of a doublet with another doublet
is decomposed as $\bm{2} \times \bm{2} = \bm{1} + \bm{1'} + \bm{2}$
and
\begin{align}
 (a \tilde{a})_{\bm{1}} = a_1 \tilde{a}_2 + a_2 \tilde{a}_1,~~~
 (a \tilde{a})_{\bm{1'}} = a_1 \tilde{a}_2 - a_2 \tilde{a}_1,~~~
 (a \tilde{a})_{\bm{2}} = \begin{pmatrix}
	      a_2 \tilde{a}_2 \\
	      a_1 \tilde{a}_1
	     \end{pmatrix},
\end{align}
while that with a triplet is $\bm{2} \times \bm{3^{(')}} = \bm{3} + \bm{3'}$
and
\begin{align}
 (a b^{(')})_{\bm{3}} = \begin{pmatrix}
	      a_1 b_2^{(')} \pm a_2 b_3^{(')} \\
	      a_1 b_3^{(')} \pm a_2 b_1^{(')} \\
	      a_1 b_1^{(')} \pm a_2 b_2^{(')}
	     \end{pmatrix},~~~
 (a b^{(')})_{\bm{3'}} = \begin{pmatrix}
	       a_1 b_2^{(')} \mp a_2 b_3^{(')} \\
	       a_1 b_3^{(')} \mp a_2 b_1^{(')} \\
	       a_1 b_1^{(')} \mp a_2 b_2^{(')}
	      \end{pmatrix},
\end{align}
where the upper (lower) sign is for $\bm{3}$ ($\bm{3'}$). The product of two triplets is 
decomposed as $\bm{3} \times \bm{3} = \bm{1} + \bm{2} + \bm{3} + \bm{3'}$ and
\begin{align}
 (b \tilde{b})_{\bm{1}} &= b_1 \tilde{b}_1 + b_2 \tilde{b}_3 + b_3 \tilde{b}_2,
 \label{eq_prod_33} \\
 (b \tilde{b})_{\bm{2}} = \begin{pmatrix}
	      b_1 \tilde{b}_3 + b_2 \tilde{b}_2 + b_3 \tilde{b}_1 \\
	      b_1 \tilde{b}_2 + b_2 \tilde{b}_1 + b_3 \tilde{b}_3
	     \end{pmatrix},~~~
 (b \tilde{b})_{\bm{3}} &= \begin{pmatrix}
	       b_2 \tilde{b}_3 - b_3 \tilde{b}_2 \\
	       b_1 \tilde{b}_2 - b_2 \tilde{b}_1 \\
	       b_3 \tilde{b}_1 - b_1 \tilde{b}_3
	      \end{pmatrix},~~~
 (b \tilde{b})_{\bm{3'}} = \begin{pmatrix}
	       2 b_1 \tilde{b}_1 - b_2 \tilde{b}_3 - b_3 \tilde{b}_2 \\
	       - b_1 \tilde{b}_2 - b_2 \tilde{b}_1 + 2 b_3 \tilde{b}_3 \\
	       - b_1 \tilde{b}_3 + 2 b_2 \tilde{b}_2 - b_3 \tilde{b}_1
	      \end{pmatrix}. \notag
\end{align}
The product $\bm{3'} \times \bm{3'}$ is decomposed in the same way and the component product 
rules can be obtained by substituting $(b, \tilde{b}) \rightarrow (b', \tilde{b}')$ in \eqref{eq_prod_33}.
Finally, the remaining non-trivial product is $\bm{3} \times \bm{3'} = \bm{1'} + \bm{2} + \bm{3} + \bm{3'}$: 
in components,
\begin{align}
 (b b')_{\bm{1'}} &= b_1 b_1' + b_2 b_3' + b_3 b_2', \\
 (b b')_{\bm{2}} = \begin{pmatrix}
	      b_1 b_3' + b_2 b_2' + b_3 b_1' \\
	      - b_1 b_2' - b_2 b_1' - b_3 b_3'
	     \end{pmatrix},~~~
 (b b')_{\bm{3}} &= \begin{pmatrix}
	       2 b_1 b_1' - b_2 b_3' - b_3 b_2' \\
	       - b_1 b_2' - b_2 b_1' + 2 b_3 b_3' \\
	       - b_1 b_3' + 2 b_2 b_2' - b_3 b_1'
	      \end{pmatrix},~~~
 (b b')_{\bm{3'}} = \begin{pmatrix}
	       b_2 b_3' - b_3 b_2' \\
	       b_1 b_2' - b_2 b_1' \\
	       b_3 b_1' - b_1 b_3'
	      \end{pmatrix}.\notag
\end{align}

\section{Diagonalization of mass matrices}
\label{sec_diag}

\subsection{Neutrino sector}

The fermion mass matrix $\mathcal M$ is in general complex and symmetric. A symmetric complex 
matrix is diagonalized by a unitary matrix $X$ in the form of $X \mathcal M X^T$ (Takagi 
diagonalization~\cite{Dreiner:2008tw}). As a concrete example, let us consider $2\times 2$ complex 
mass matrix
\begin{align}
	\mathcal M =  \begin{pmatrix} B & C \\ C & D \end{pmatrix}.
\end{align}
A general unitary matrix may be expressed as
\begin{align}
	X = \begin{pmatrix} \cos\theta & e^{i\eta} \sin\theta \\ -e^{-i\eta} \sin\theta & \cos\theta \end{pmatrix}
            \begin{pmatrix} e^{i\alpha} & 0 \\ 0 & e^{i\beta} \end{pmatrix},
\end{align}
where $\theta$, $\eta$, $\alpha$ and $\beta$ are real parameters. We find that $X^T \mathcal M X$ 
becomes diagonal if we take
\begin{align}
	\tan2\theta = \frac{2|C^*D + C B^*|}{|D|^2-|B|^2},~~~~~~e^{i\eta} = \frac{C^*D + C B^*}{|C^*D + C B^*|}.   
\label{tan2theta}
\end{align}
The parameters $\alpha$ and $\beta$ can be fixed if one wants to make the mass eigenvalues real.
The neutrino mass matrix in TM$_1$ model (\ref{UmU}) can be diagonalized using this expression 
by identifying $u_{23} = X$.

\subsection{Flavon sector}

There are only two types of mass matrices that appear in the analysis of the scalar sector of flavon fields. 
In this appendix, we summarize all of their eigenvectors and eigenvalues.  First, in Sec.~\ref{sec:phiell} 
and \ref{sec:phi2}, we obtain the mass matrices of the form of
\begin{align}
 \mathcal{M}_2^2 = \left(
 \begin{array}{cc}
  B & C \\
  C & B \\
 \end{array}
 \right),\label{eq_m2sq}
\end{align}
which is diagonalized by an orthogonal matrix $O$ as
\begin{align}
 O \mathcal{M}_2^2 O^T =
 \left(
 \begin{array}{cc}
  B-C & 0 \\
  0 & B+C \\
 \end{array}
 \right)~~;~~
 O = \frac{1}{\sqrt{2}} \left(
 \begin{array}{cc}
  1 & -1 \\
  1 & 1 \\
 \end{array}
 \right).
\end{align}
Note that there exists a massless field if $B = \pm C$. For the analysis of triplet flavon fields in 
Sec.~\ref{sec:phi3} and \ref{sec:psi3}, we obtain the mass matrices of the form of
\begin{align}
 \mathcal{M}_3^2 = \left(
 \begin{array}{ccc}
  D & E & E \\
  E & F & G \\
  E & G & F \\
 \end{array}
 \right),\label{eq_m3sq}
\end{align}
which can be diagonalized as
\begin{align}
 V \mathcal{M}_3^2 V^T &=
 \left(
  \begin{array}{ccc}
   F-G & 0 & 0 \\
   0 & \frac{1}{2} (D+F+G-\kappa) & 0 \\
   0 & 0 & \frac{1}{2} (D+F+G+\kappa) \\
  \end{array}
 \right),\\
 V &= \left(
 \begin{array}{ccc}
  0 & -\frac{1}{\sqrt{2}} & \frac{1}{\sqrt{2}} \\
  \frac{D-F-G-\kappa}{E N_{-}} & \frac{2}{N_{-}} & \frac{2}{N_{-}} \\
  \frac{D-F-G+\kappa}{E N_{+}} & \frac{2}{N_{+}} & \frac{2}{N_{+}} \\
 \end{array}
 \right),
\end{align}
for $E\neq 0$ where
\begin{align}
 \kappa \equiv \sqrt{(D-F-G)^2+8E^2},
\end{align}
and $N_{\pm}$ are proper normalization factors with which the squared sum of each line of $V$ becomes 
one. Note that there is a massless mode if $F=G$, as seen in Sec.~\ref{sec:phi3} and \ref{sec:psi3}.



\end{document}